\documentclass[]{aa}  

\usepackage{graphicx}
\usepackage[varg]{txfonts}
\usepackage{natbib}
\bibpunct{(}{)}{;}{a}{}{,} 

\begin{document}
   \title{Dark influences II: gas and star formation in minor mergers of dwarf galaxies with dark satellites}

   \subtitle{}

  \author{T. K. Starkenburg\inst{1} 
         \and A. Helmi\inst{1} 
         \and L. V. Sales\inst{2,3}
         }

  \institute{Kapteyn Astronomical Institute, University of Groningen,
             P.O. Box 800, 9700 AV Groningen, The Netherlands\\
             \email{tjitske@astro.rug.nl}
             \and Harvard-Smithsonian Center for Astronomy, 60 Garden Street, Cambridge, MA 02138, USA
             \and Department of Physics and Astronomy, University of California, Riverside, CA 92521, USA  
             }

  \date{Received date / Accepted date }

\abstract{It has been proposed that mergers induce starbursts and lead to important morphological changes in galaxies.  Most studies so far have focused on large galaxies, but dwarfs might also experience such events, since  the halo mass function is scale-free in the concordance cosmological model. Notably, because of their low mass, most of their interactions will be with dark satellites.}
{In this paper we follow the evolution of gas-rich disky dwarf galaxies as they experience a minor merger with a dark satellite. We aim to characterize the effects of such an interaction on the dwarf’s star formation, morphology, and kinematical properties.}
{We performed a suite of carefully set-up hydrodynamical simulations of dwarf galaxies that include dark matter, gas, and stars merging with a satellite consisting solely of dark matter. For the host system we vary the gas fraction, disk size and thickness, halo mass, and concentration, while we explore different masses, concentrations, and orbits for the satellite.}
{We find that the interactions cause strong starbursts of both short and long duration in the dwarfs. Their star formation rates increase by factors of a few to $10$ or more. They are strongest for systems with extended gas disks and high gas fractions merging with a high-concentration satellite on a planar, radial orbit. In contrast to analogous simulations of Milky Way-mass galaxies, many of the systems experience strong morphological changes and become spheroidal even in the presence of significant amounts of gas.}
{The simulated systems compare remarkably well with the observational properties of a large selection of irregular dwarf galaxies and blue compact dwarfs. This implies that mergers with dark satellites might well be happening but not be fully evident, and may thus play a role in the diversity of the dwarf galaxy population.}


\titlerunning{Gas and star formation in minor mergers of dwarf galaxies with dark satellites}
\maketitle
\section{Introduction}

In the lambda cold dark matter ($\Lambda$CDM) paradigm, small dark
matter halos are abundant. Most of these halos, with
$M_{\mathrm{vir}} < 10^9 M_{\sun}$, are predicted to be strongly affected by
reionization, photo-evaporation, and/or supernova feedback
(\citealp{Gnedin2000, Hoeftetal2006, Kaufmannetal2007, Okamotoetal2008,
  Gnedinetal2009, Lietal2010, Sawalaetal2013}; \citealp[but see
also][]{TaylorWebster2005, Warrenetal2007}). These processes thus cause
progressively larger numbers of small dark matter halos to be almost
completely dark. The existence of such dark galaxies is a solution that is often
suggested to the missing satellites problem
\citep{Klypinetal1999, Mooreetal1999}.

Dwarf galaxies are known to be very inefficient at forming stars
\citep{Blantonetal2001, RobertsonKravtsov2008}, to have very low
baryon fractions \citep{Gnedin2000, Hoeftetal2006, Crainetal2007}, and to
generally be gas-rich if in the field. This is consistent with the
expectation that the stellar-to-halo-mass ratio must decrease steeply
toward lower masses \citep{Behroozietal2013, Mosteretal2013,
  KormendyFreeman2014, Garrison-Kimmeletal2014, Sawalaetal2015}. On
the other hand, the halo mass function is predicted to be almost
completely scale-free \citep{vdBoschetal2005,vdBoschJiang2014},
i.e., similar for field dwarf and large disk galaxies. However, the
subhalos of dwarf galaxies must have much lower baryonic component masses,
and so most of their satellites will be dark \citep{Helmietal2012}.

Mergers of gas-rich galaxies are often thought to give rise to
bursts of star formation \citep{MihosHernquist1994a,MihosHernquist1994b,Teyssieretal2010,
  Bournaudetal2011}, although simulations suggest this depends 
on the merger mass ratio \citep[e.g.,][]{DiMatteoetal2007}. From the observational perspective, \citep[see e.g.,][]{Ellisonetal2011, Willettetal2015}, even minor mergers have been shown to significantly contribute to local star formation \citep{Kaviraj2014b, Kaviraj2014a, Willettetal2015}. In the case of dwarf
galaxies it has been suggested that interactions are responsible for the class of blue compact
dwarfs (BCDs) \citep{Paudeletal2015}: dwarf galaxies with a significant, centrally
concentrated young stellar population \citep[e.g.,][]{dePazetal2003}. In
general many dwarf systems with increased star formation rates have
been found to be irregular or to show signs of disturbances
\citep{Tayloretal1995,EktaChengalur2010,Lopez-Sanchez2010,Holwerdaetal2013,Lellietal2014c,KnapenCisternas2015}, but in a number of cases no visible companion
has been found \citep{Broschetal2004, EktaChengalur2010,Lopez-Sanchez2010, Lellietal2014c}. Other possible
origins of the increase in star formation and the irregular morphology
are cosmological gas inflows \citep[see for
example][]{Verbekeetal2014} or re-accretion of material blown out by
previous starbursts, and varying internal instabilities \citep[e.g.,][]{Meureretal1998,vanZeeetal2001,Lellietal2014a,Elmegreenetal2012,BekkiFreeman2002}.

In view of the above discussion, it appears plausible that some of
these starbursts could be induced by interactions with dark
satellites. Following the first paper by \citet{SH15} here we focus on
the effect of a merger on a gas rich dwarf galaxy. \citet{SH15} ran a
suite of collissionless simulations of dwarf galaxies and their dark
satellites, and showed that these could severely alter the morphology
and kinematics of the dwarf. In this paper we extend these simulations
to dwarf galaxies with varying gas fractions and study the effects of
the merger on the star formation rates, gas and stellar morphology and
kinematics. The initial conditions of the dwarf galaxies and their
satellites and the parameters for the simulations are described in
Sect.~\ref{Models}. We report our results and their dependence on
properties of the systems and the interactions in
Sect.~\ref{Results}. We compare these SPH-results with the
collissionless simulations from \citet{SH15} in
Sect.~\ref{Comparisoncoll} and to observational results in
Sect.~\ref{Comparisonobs}. We conclude by discussing our results in
Sects.~\ref{Discussion} and give a summary of our main findings in
Sect.~\ref{Conclusions}.

\section{Models}
\label{Models}

We perform a suite of controlled simulations of isolated dwarf galaxies and mergers with their (dark) satellites. The simulations are run using the OWLS version \citep[last described in ][]{Schayeetal2010} of the N-body/SPH-code Gadget-3 \citep[based on][]{SpYW01, Sp05} with implementations for star formation and feedback as described in \citet{SDV08, DVS08}. 

\subsection{Initial conditions}
The setup of the initial conditions is based on \citet{SDMH05, SDV08, DVS08} and a more complete description can be found in \citet{SH15}. Here we briefly describe the
initial structure of the dwarf galaxy and of the satellite. The values of the structural and orbital parameters for all the simulations are listed in Tables~\ref{hostparm} and~\ref{satparm}.

\subsubsection{Main (disky dwarf) galaxies}

The host dwarf galaxies have several components including a dark halo, a stellar disk and a gaseous disk. The dark matter halo follows a Hernquist profile \citep{Hernquist1990},
\begin{equation}
\rho(r) = \frac{\rho_0}{(r/a)(1+r/a)^3}
,\end{equation}
where the parameters $\rho_0$ and $a$ are set by an equivalent NFW \citep{NFW} profile, such that the total mass of the Hernquist halo equals the virial mass of the NFW halo and their profiles have similar inner densities ($r_s\rho_{0\mathrm{,NFW}} = a\rho_{0\mathrm{,H}}$) \citep{SDMH05}.

We consider three different concentrations $c_{\rm host} = r_{\rm vir}/r_{s_{\rm NFW}}$ for the
halos: 5, 9 and 15. The latter two are consistent with the
mass~--~concentration relation found in large cosmological simulations
\citep{Maccioetal2008, MunozCuartasetal2011} for the mass-scale
considered here. The lowest $c_{\rm host}$ corresponds to the best-fit
NFW model for the Fornax dwarf spheroidal galaxy from 
\citet{BreddelsHelmi2013}. The velocities of the halo particles are
set using the distribution function of a Hernquist halo. Since the
contribution of the disk is hereby initially neglected, the
halo shows a slight adiabatic contraction at the start of the
simulations, which stabilizes within approximately 0.5 Gyr.

The stellar masses of the systems follow generally (extrapolated)
stellar mass~--~halo mass relations available in the literature
\citep{Behroozietal2013,Mosteretal2013,Garrison-Kimmeletal2014,Sawalaetal2015}. The
gas fraction, $f_g =M_{\mathrm{gas}}/(M_{\star} + M_{\mathrm{gas}})$,
ranges from $f_g = 0.3$ to $f_g = 0.9$ in agreement with observational estimates \citep{Huangetal2012, McQuinnetal2015}. This means that the
baryonic mass can be quite high, and that in the most gas-rich systems $M_{\mathrm{gas}} = 9 M_{\star}$.

Both the stellar and gaseous disk follow an exponential surface
density profile with radius. The scale lengths of the stellar disks
are close to the values expected from \citet{MoMaoWhite1998, SpW99}
for the $M_{\mathrm{vir}} = 1.4 \times 10^{10} M_{\sun}$ and $c_{\rm host}=15$
systems, and also for the lowest mass dwarf ($M_{\mathrm{vir}} = 0.55
\times 10^{10} M_{\sun}$) with $c_{\rm host}=5$ dark matter halo, assuming that the disk angular momentum fration equals the disk mass fraction ($j_d = m_d$). For completeness we explore for 
two systems ($M_{\mathrm{vir}}=0.97 \times 10^{10}
M_{\sun}$ and $M_{\mathrm{vir}}=1.4 \times 10^{10} M_{\sun}$) smaller disk scale lengths. For all disks the exponential
radial profiles and the scale lengths are stable during evolution in
isolation. The stellar disk is further described by an isothermal
vertical distribution with a constant scale height such that $0.1 R_d
< z_0 < 0.5 R_d$:
\begin{equation}
\label{diskdensprof}
\rho_{d,\star} (R,z)=\frac{M_{d,\star}}{4 \pi R_d^2 z_0} \ \mathrm{ exp}\left(-\frac{R}{R_d}\right) \ \mathrm{ sech}^2 \left(\frac{z}{z_0}\right). 
\end{equation}
We consider thicker disks for the lowest mass simulated galaxies. This is in concordance with observations \citep{YoachimDalcanton2006, Sanchez-Janssenetal2010} and also expected since gas cooling is less efficient in smaller halos \citep{Kaufmannetal2007, RobertsonKravtsov2008}.

We explore two different scale lengths of the gaseous disk in comparison to the stellar disk, namely $R_g = R_d$, and $R_g=2R_d$. The vertical distribution of the gas is determined by requiring hydrostatic equilibrium,
\begin{equation}
-\frac{1}{\rho_g}\frac{\partial P}{\partial z} - \frac{\partial \Phi}{\partial z}=0
.\end{equation}
Using an effective equation of state (EOS), with $\gamma_{\mathrm{EOS,eff}}=\left( d \textrm{ ln } P\right)/\left( d \textrm{ ln } \rho\right)$, this can be written as 
\begin{equation}
  \frac{\partial \rho_g}{\partial z} = - \frac{\rho_g^2}{\gamma_{\mathrm{EOS,eff}} P} \frac{\partial \Phi}{\partial z}
\label{eq:gamma}
  .\end{equation}
In our simulations we use the effective equation of state of the multiphase ISM model by \citet{SDV08, DVS08}. The gas is governed by the effective equation of state when the density of the gas is above the threshold for star formation as set by the subgrid physics discussed in Sect.~\ref{SPH}, while at lower densities the gas is assumed to follow an isothermal equation of state. 

The vertical distribution of the gas can be iteratively determined as a function of radius using a fine logarithmic grid in the $R$--$z$ plane \citep{SDMH05}, where the surface density of the gas, $\Sigma_{d,\mathrm{gas}} \left(R\right)$, is set by the exponential radial profile and the chosen scale length,
\begin{equation}
\Sigma_{d,\mathrm{gas}} \left(R\right) = \frac{M_{d,\mathrm{gas}}}{2 \pi R_g^2} \ \mathrm{ exp}\left(-\frac{R}{R_g}\right) = \int \rho_g \left(R,z\right) \mathrm{d}z.
\end{equation}
To calculate the potential in Eq.~(\ref{eq:gamma}) we follow \citet{SDMH05} in using a tree code and a discretized mass distribution to represent the disk components, adding contribution of the dark matter halo analytically, and use a grid of $4096 \times 64 \times 128$ in the radial, vertical and azimuthal directions.

In \citet{SH15} we adapted the way in which the velocity structure of the stellar disks is set up. Contrary to large galaxies, for lower mass systems the epicyclic approximation breaks down in a large part of the disk which, if used, can cause a non-physical streaming velocity. Therefore we set up the velocity structure only assuming the epicyclic approximation in the part of the disk where it is still valid and fit a smooth quadratical curve to the rotation curve closer in. Moreover, the potential of the model dwarf galaxies is dominated by their dark matter halo at all radii, so we assume that the velocity ellipsoid is aligned with the spherical coordinate system instead of the cylindrical coordinate system of the disks.  For the gas, the initial azimuthal streaming velocity is determined through the gravitional, pressure, and centrifugal forces:
\begin{equation}
 v^2_{\phi,\mathrm{gas}} = R \left( \frac{\partial \Phi}{\partial R} + \frac{1}{\rho_g} \frac{\partial P}{\partial R} \right)
.\end{equation}

\subsubsection{Satellites}

The satellite consists solely of a dark matter halo with a mass of
$20\%$ of the dwarf galaxy's halo mass in most of our simulations. Its concentration is
determined by following the mass~--~concentration relation from
\citet{MunozCuartasetal2011}. In Sect.~\ref{satmass} we explore the
effect of a two or four times smaller mass and a higher concentration
for the satellite.

Most of the satellite orbits we consider are close to completely radial with initially none or a small (prograde) tangential velocity, although we also explore more circular orbits (see Table \ref{satparm} and Sect. \ref{orbit}). At the start of the simulations the satellite is placed at a
distance of $0.67 r_{\mathrm{vir}}$ from the center of the host with a radial velocity that is small with respect to the local circular velocity (so the satellite is initially close to apocenter). The orbit is either in the plane of the host disk or has an inclination of 30 (or in one case 60) degrees, similar to other studies of disk thickening \citep{VH08, Mosteretal2010}.

 \begin{table}[0.9\textwidth]
   \caption{\label{hostparm} Structural parameters for the host dwarf galaxies.}
         $$ 
         \begin{array}{llllllllr}
            \hline
            \noalign{\smallskip}
              \mathrm{Model} & M_{\rm vir} & r_{\mathrm{vir}} & c & M_{\star} & R_d & \displaystyle\frac{z_0}{R_d} & f_g\\
               & 10^{10} M_{\sun} & \mathrm{kpc} &  & 10^8 M_{\sun} & \mathrm{kpc} &  & \\           
            \noalign{\smallskip}
            \hline
            \hline
            \noalign{\smallskip}
            \noalign{\smallskip}
            {\rm A} & 5.6 & 77 & 9 & 1.4 & 0.93 & 0.1 & 0.5 \\
            {\rm B} & 2.2 & 56 & 15 & 0.27 & 0.78 & 0.2 & 0.75 \\
            {\rm C1} & 1.4 & 48 & 15 & 0.11 & 0.78 & 0.3 & 0.9 \\
            {\rm C2} &\ldots & \ldots & \ldots & \ldots & 0.39 & \ldots & \ldots \\
            {\rm C3} &\ldots & \ldots & 5 & \ldots & 0.78 & \ldots & \ldots \\
            {\rm C4} & 1.4\tablefootmark{a} & \ldots & 15 & 0.55 & \ldots & 0.2 & 0.5 \\
            {\rm C5} &\ldots\tablefootmark{a} & \ldots & \ldots & 0.77 & \ldots & \ldots & 0.3 \\
            {\rm D1} & 0.97 & 42 & 5 & 0.044 & 0.95 & 0.3 & 0.9 \\
            {\rm D2} & \ldots & \ldots & 15 & \ldots & \dots & \ldots & \ldots  \\
            {\rm D3} & \ldots & \ldots & 5 & \ldots & \ldots & 0.5 & \ldots  \\
            {\rm D4} & \ldots & \ldots & 5 & \ldots & 0.48 & 0.3 & \ldots  \\
            {\rm D5} & \ldots & \ldots & 5 & \ldots & \ldots & 0.5 & \ldots  \\
            {\rm E1} & 0.55\tablefootmark{b} & 27 & 5 & 0.22 & 0.95 & \ldots & 0.5 \\
            {\rm E2} & \ldots\tablefootmark{b} & \ldots & \dots & 0.31 & \ldots & \ldots & 0.3 \\
            \noalign{\smallskip}
            \hline
            \hline
         \end{array}
     $$ 
     \tablefoot{
     $\ldots$ denotes that the value is equal to that reported in the row above. \\
     \tablefoottext{a}{These systems are equivalent to the \emph{disk3}-systems from \citet{SH15} but include gas.}
     \tablefoottext{b}{These systems are equivalent to the \emph{FNX-analog}-systems from \citet{SH15} but include gas.}
     }
   \end{table}

 \begin{table}[\textwidth]
      \caption{\label{satparm} Parameters for the satellites and their orbits}
         $$ 
         \begin{array}{llllrcr}
            \hline
            \noalign{\smallskip}
              \displaystyle M_{\rm sat}/M_{\rm vir_{\rm main}} & c_{\mathrm{sat}} & \displaystyle\frac{v_r}{v_{\mathrm{vir_{\rm main}}}} & \displaystyle\frac{v_t}{v_{\mathrm{vir_{\rm main}}}} & r_{\mathrm{apo}}/r_{\mathrm{peri}}\tablefootmark{a} & \mathrm{inclination} \\    
            \noalign{\smallskip}
            \hline
            \hline
            \noalign{\smallskip}
            \noalign{\smallskip}
             0.2 & 15 & -0.08 & 0.06 &  \sim40  & 30 \\
             \ldots & \ldots & \ldots & \ldots &  \sim40  & 60 \\
             \ldots & \ldots & 0 & \ldots & \sim40 & 0  \\
             \ldots & 25 & -0.08 & \ldots & \sim40 & 30 \\
             \ldots & \ldots & 0 & \ldots & \sim35 & 0 \\
             \ldots & \ldots & \ldots & 0.86 & 2 & \ldots \\
             \ldots & \ldots & \ldots & 0.5 & 6 & \ldots \\
             0.1 & 16 & 0 & 0.06 & \sim45 & 0 \\
             \ldots & 25 & \ldots & \ldots & \sim45 & \ldots \\
             0.05 & 17 & \ldots & \ldots & \sim40 & \ldots \\
             \ldots & 25 & \ldots & \ldots & \sim35 & \ldots \\ 
            \noalign{\smallskip}
            \hline
            \hline
         \end{array}
     $$ 
     \tablefoot{
      $\ldots$ denotes that the value is equal to that reported in the row above. \\
       \tablefoottext{a}{The apo-to-peri ratio is defined for the first pericentric passage. For very radial orbits, this ratio is uncertain (by $\sim 20\%$) because of the dependency on the time-sampling around this passage.
       }
     }   \end{table}

\subsection{Numerical parameters}
\label{NumParam}
We consider two different numerical setups for the host dwarf
galaxies. For the \emph{disk3-gas} (models C4 and C5) and \emph{FNX-gas} (models E1 and E2) systems that we
compare to the collissionless simulations of \citet{SH15}, we use $5
\times 10^6$ particles to represent the dwarf's dark matter halo
and $10^5$ particles for the gaseous and for the stellar disk irrespective
of the value of $f_g$. For all other hosts the $2 \times 10^5$
baryonic particles are divided among gas and stars according to $f_g$
so that all gas and star particles have initially equal mass, and the
dark halo is represented by $10^6$ particles.

The satellite is in all cases represented by $5 \times 10^5$
particles which gives a dark matter particle mass for the satellite of $2 \times 10^3\ M_{\sun} \lesssim m_{\mathrm{sat}} \lesssim 2 \times 10^4\ M_{\sun}$. The softening lengths used are $0.025$ kpc and $0.016$ kpc for the
host halo and satellite respectively, and $0.008$ kpc for both the gas
and stars. These values are chosen following \citet{VH08} and
\citet{Athanassoulaetal2000} and produce stable systems. All
numerical parameters are summarized in Table~\ref{numparm}.
 \begin{table}[0.7\textwidth]
      \caption{\label{numparm} Numerical parameters}
        $$ 
         \begin{array}{llllllllllr}
            \hline
            \noalign{\smallskip}
             {\rm Model } & \displaystyle \frac{M_{\rm sat}}{M_{\rm vir_{\rm main}}} & N_{\mathrm{DM}} & N_{\mathrm{gas}} & N_{\star} & m_{\mathrm{DM}} & m_{\mathrm{bar}}  \\
               & & \times 10^6  & \times 10^5 & \times 10^5 & 10^4 M_{\sun} & 10^2   M_{\sun}  \\           
            \noalign{\smallskip}
            \hline
            \hline
            \noalign{\smallskip}
            \noalign{\smallskip}
            {\rm A} &  0.2 & 1 &  1 & 1 & 5.60 & 13.8 \\
            \ldots &  0.1 & \ldots & \ldots & \ldots &  \ldots & \ldots  \\
            \ldots &  0.05 & \ldots & \ldots & \ldots & \ldots & \ldots  \\
            {\rm B}  & 0.2 & \ldots& 1.5 & 0.5  & 2.12  & 5.34 \\
            {\rm C1-C3}  & \ldots & \ldots & 1.8  & 0.2  & 1.45  & 5.55 \\
            {\rm D}  & \ldots & \ldots & \ldots & \ldots & 0.97  & 2.21 \\
            {\rm C4}  & \ldots & 5  & 1 & 1 & 0.27 & 5.51  \\
            {\rm C5}  & \ldots & \ldots  & \ldots & \ldots & \ldots & 7.71\tablefootmark{a}  \\
            {\rm E1}  & \ldots & \ldots  &  \ldots & \ldots & 0.17  & 3.35\\
            {\rm E2}  & \ldots & \ldots  &  \ldots & \ldots & \ldots  & 4.69\tablefootmark{b} \\
            \noalign{\smallskip}
            \hline
            \hline
         \end{array}
     $$ 
\tablefoot{
       $\ldots$ denotes that the value is equal to that reported in the row above. \\
       \tablefoottext{a}{For the system C5 $m_{\star} \neq m_{\mathrm{gas}}$ and the value reported gives $m_\star$. $m_{\mathrm{gas}} = 3.31 \times 10^2\ M_{\sun}$.}
       \tablefoottext{b}{For the system E2 $m_{\star} \neq m_{\mathrm{gas}}$ and the value reported gives $m_\star$. $m_{\mathrm{gas}} = 2.01 \times 10^2\ M_{\sun}$.}
       }
   \end{table}

\subsection{Star formation and feedback prescription}
\label{SPH}

We use the star formation and stellar feedback prescription from
\citet{SDV08} and \citet{DVS08}. The star formation prescription is based on
empirical laws and has very good numerical properties. \citet{SDV08}
have shown how the gas surface densities in the 
Kennicutt-Schmidt law \citep{Kennicutt1998} can be related to pressure by assuming that the scale height of the gas disk is of
order of the local Jeans scale and that the gas is in local
hydrostatic equilibrium and self-gravitating. This is combined with a
polytropic effective equation of state for the multiphase interstellar
medium with a slope of $\gamma_{\rm EOF}=4/3$, which always ensures a constant Jeans mass (so independent of the local gas
density). We follow \citet{Schaye2004} in using a density threshold $n_H=0.1\ \mathrm{cm}^{-3}$ which corresponds to a surface density threshold of $\sim10\ M_{\sun}\ \mathrm{pc}^{-2}$ and a temperature threshold of $10^4$~K \citep{SDV08}. We do include radiative cooling but no metals and therefore no metal-line cooling or chemical enrichment.

The feedback model is described in \citet{DVS08} and consists of kinetic supernova winds. The model is governed by the wind speed $v_w$, and mass loading $\eta = \dot{M}_w/\dot{M}_{\star}$, which are  related to the fraction of the kinetic energy injected by supernovae (SNe) per solar mass, $\epsilon_{\mathrm{SN}}$, also called the feedback efficiency as
\begin{equation}
f_w = \frac{\eta v_w^2}{2 \epsilon_{\mathrm{SN}}}.
\end{equation}
Following \citet{DVS08} we set $\epsilon_{\mathrm{SN}} = 1.8 \times
10^{49}\ \mathrm{erg}\ M_{\sun}^{-1}$ which is appropriate for a
\citet{Chabrier2003} initial mass function, a stellar mass range
$0.1$~--~$100\ M_{\sun}$ and all stars above $6\ M_{\sun}$ ending as
core-collapse SNe. The wind particles remain as such for $t_w = 1.5
\times 10^7\ \mathrm{yr}$ and during that time are not able to be
kicked again by another SN or participate in star formation, but they
are not decoupled hydrodynamically.

Both observational data \citep{SchwartzMartin2004, Martin2005} and
theoretical models \citep{Okamotoetal2010, Lagosetal2013} suggest that
wind velocities are lower and the mass loading rates are higher for
low mass galaxies and low star formation rates. Therefore, rather than
making a single choice for the numerical parameters, we experimented with a
selection of different parameters values as described in the next section.

\begin{figure}[b]
\includegraphics[width=0.5\textwidth]{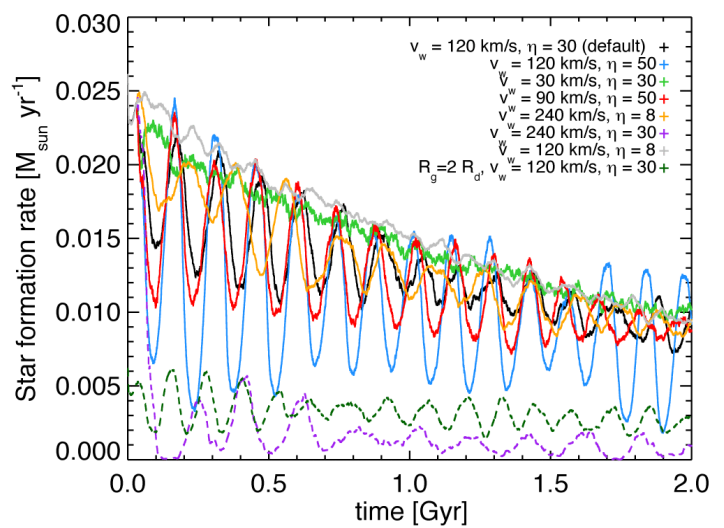}
\caption{\label{SFRtests} Star formation rates for model C1
  ($M_{\mathrm{vir}}=1.4 \times 10^{10} M_{\sun}$) with different values of the 
  feedback parameters: wind velocity and mass
  loading depicted in different colors as indicated by the inset for $R_g = R_d$, except for dark green where $R_g =
  2 R_d$.}
\end{figure}

\subsection{Evolution in isolation}
\label{isolation}

\subsubsection{Exploration of subgrid parameters}
\label{subgridexploration}

To identify a reasonable default model for the dwarf galaxies we
explored a small set of combinations of the large parameter space
and varied two of the stellar feedback parameters, namely the wind speed $v_w$ ($30$, $90$, $120$, $240$, and $600$ km/s) and the mass loading $\eta$ (2, 8, 10, 20,
30, 40, 50).

Our goal is for a galaxy in isolation to have a fairly continuous
star formation rate on a timescale comparable to that of the merger. This is important to be able to pin down to the effect of
the merger. However since we do not model fresh gaseous infall all
star formation rates will decline due to gas depletion. We also pay
attention to ensure that the gas is not all converted into stars or
blown away by stellar feedback too quickly.  Nonetheless, a tight
correlation exists between the initial gas mass and its distribution
and the early star formation rate. Via the threshold for star formation, this also depends on the
gravitational potential (and hence on the mass and concentration of
the dark matter halo and the mass, scale length, and scale height of
the stellar disk).

\begin{figure*}[!ht] 
\includegraphics[angle=270, width=\textwidth]{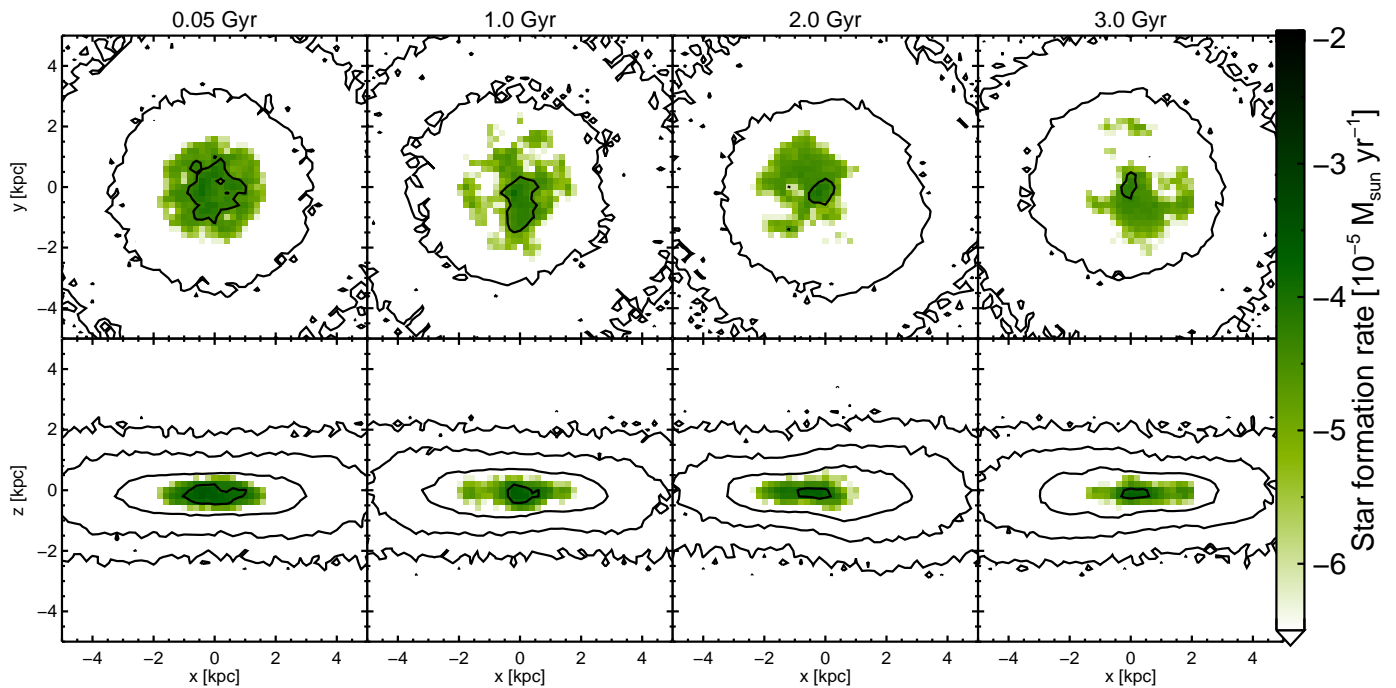}
\caption{\label{iso} Face-on (top) and edge-on (bottom) density contours of the gas in the disk of model A with $R_g=2 R_d$, evolved in isolation. The contour levels are at $0.25$, $1$, $4$, and $16 \times 10^{20}\ \mathrm{N}\ \mathrm{cm}^{-2}$. The local star formation regions (and their amplitude) are shown in green at different moments during the simulation.}
\end{figure*}

The star formation rate over a period of 2 Gyr for the dwarf
galaxy model C1 ($M_{\rm vir}=1.4 \times 10^{10} M_{\sun}$,
$f{_\mathrm{gas}}=0.9$, $R_g = R_d$, and $z_0 = 0.3 R_d$), for
varying feedback parameters is shown in Fig.~\ref{SFRtests}. The star
formation rates for all prescriptions are within a factor of a few of
each other and in all cases star formation is sustained over a long
timescale. For the more extended gas disk ($R_g=2R_d$, dashed dark green) the star formation rates are
about a quarter of the $R_g=R_d$-case (black). For the most efficient
feedback ($v_w = 240 \mathrm{km/s}$ and $\eta = 30$ so
$E_{\mathrm{wind}}/E_{\mathrm{SN}}=96\%$, dashed purple) the gas disk is blown apart
right after the onset of star formation and subsequent star formation
occurs at a similar rate as for the more extended gas disk. The
periodicity seen in the first $\sim 1$ Gyr is driven by the
fact that star formation starts in the whole disk at the same
time. This is enhanced if the disk is not perfectly centered in the
potential but is reduced at later times and for lower star formation rates.
The amplitude of the oscillations in star formation also depends on the
star formation and feedback parameters. For example the amplitude of
the periodicity is larger for higher mass loading of the wind (compare
for example the gray, black, and blue star formation rates in
Fig.~\ref{SFRtests}). Reassuringly, the evolution is very similar for
settings with comparable feedback efficiencies (see for example the
red, yellow, and black curves in Fig.~\ref{SFRtests} all with
$E_{\mathrm{wind}}/E_{\mathrm{SN}}\approx 24\%$).

We tested the influence of decoupling the wind particles from
star formation by comparing with a run without decoupling, and found
that this does not change the star formation rate and evolution of the
gas disk. We further explored the effects of using higher values for
the density threshold $n_H$ (1 and 10 cm$^{-3}$), as proposed in the recent literature \citep[e.g.,][]{Governatoetal2010}, but found that this
restricts star formation to the very center of the galaxy and does not enhance
the fragmentation of the disk gas as we do not include metal-line cooling, and therefore does not lead to a more realistic system.

As default model we take the star formation and feedback parameters corresponding to the black curve in Fig.~\ref{SFRtests} for $R_g=R_d$ (dark green for $R_g=2R_d$). The default wind velocity is $v_w = 120$ km/s with a mass loading of $\eta = 30$, resulting in $f_w = 24\%$.

\begin{figure}[!h] 
\includegraphics[width=0.42\textwidth]{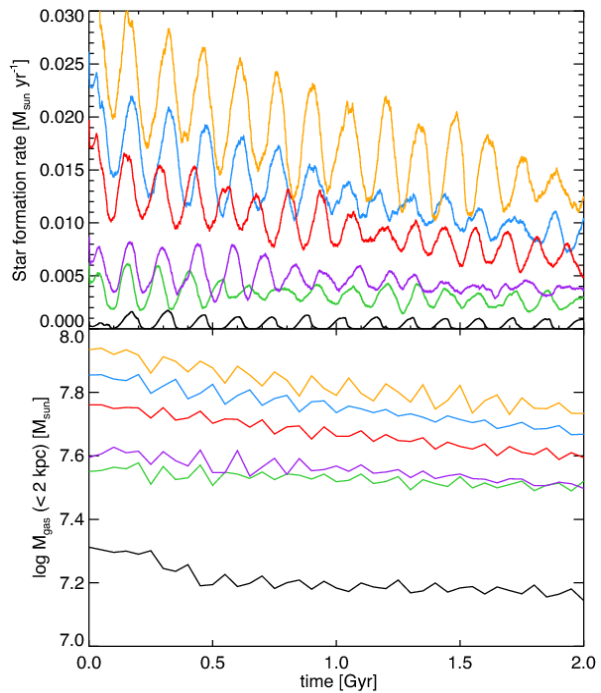}
\caption{\label{SFRgsmiso} Star formation rates (top panel) and
  central gas masses (bottom panel) for all systems in isolation with
  $R_g = R_d$ for models A (orange), B (red), C1 (blue), D1 (black),
  and with $R_g = 2 R_d$ for models A (purple) and C1 (green). Note that although model C1 (blue) 
  is less massive than B (red), the
  star formation rate is higher for C1 because it has a
  higher gas mass in the center due to its higher initial gas fraction.}
\end{figure}

\begin{figure*} 
\includegraphics[width=\textwidth]{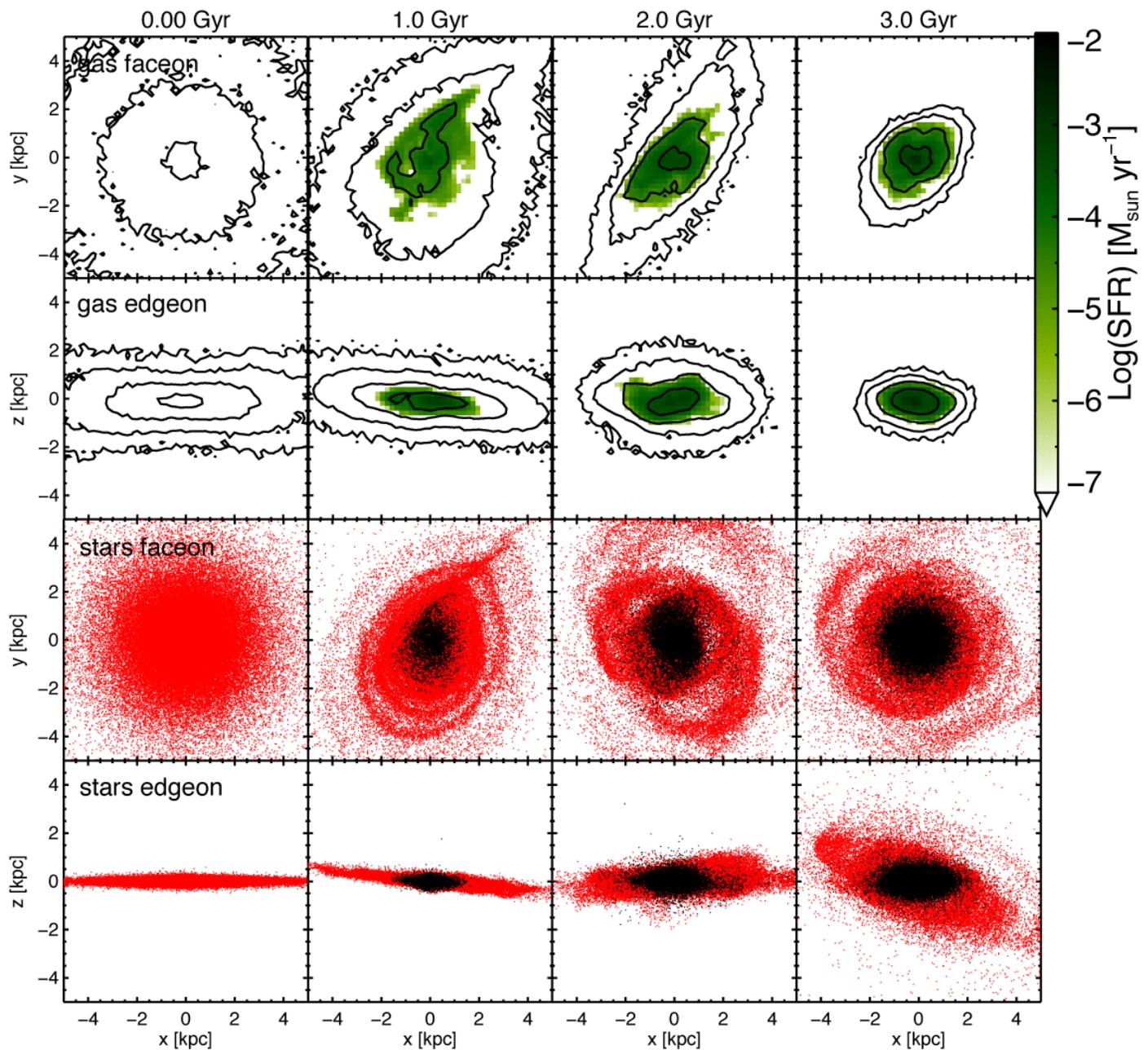}
\caption{\label{merger} Evolution of model A ($M_{\star}=1.4 \times
  10^8 M_{\sun}$) with $R_g=2 R_d$ merging with a 20\% mass satellite on a
  co-planar, very radial orbit. The top rows show the face-on and
  edge-on, respectively, view of the gas in the disk (contours at
  $0.25$, $1$, $4$, and $16 \times 10^{20}\ \mathrm{N}\ \mathrm{cm}^{-2}$) with
  the gas that is currently forming stars highlighted in green (see
  colorbar for relative values). The bottom panels show the old stellar
  component in red, and newly formed star particles in black, at
  different times during the merger.}
\end{figure*}

\subsubsection{The default model}
\label{default}

Figure~\ref{iso} shows snapshots of the gas distribution and state for
one of the most massive dwarf galaxies, model A
($M_{\mathrm{vir}}=5.6\times 10^{10} M_{\sun}$ and $M_{\star} = 1.4
\times 10^{8} M_{\sun}$), evolved in isolation. For all plots in this
paper the plane of the disk is defined as the plane perpendicular to
the angular momentum vector of the inner 50\% of the initial stellar
disk particles. As mentioned above the stellar feedback parameters are
set such that the star formation rate and the disk of the dwarf galaxy
are reasonably stable for several Gyr. Since the gas in the outskirts
of the disk is not dense enough, most of the star
formation takes place within the central~$\sim 2\, \mathrm{kpc} \, (\sim
R_g = 2R_d$ for this model). Due to the stellar feedback which blows
gas away that eventually falls back again, the star formation is
patchy and locally bursty.

The star formation rates (SFR) for a subset of the dwarf galaxies run in isolation using
our default star formation and feedback parameters are plotted in the top panel of
Fig.~\ref{SFRgsmiso}. The average amplitude of the star formation rate
clearly depends on the total mass of the galaxy, the amount of gas in
the disk and its initial extent. Indeed, for a more
extended disk set up in equilibrium, a smaller fraction of gas will be
above the density threshold for star formation.

The amount of gas in the central part of the disk is shown in the
bottom panel of Fig.~\ref{SFRgsmiso}. That both the gas fraction and
the mass of the dwarf galaxy are important is made explicit by
comparing model B ($M_{\mathrm{vir}} = 2.2 \times 10^{10} M_{\sun}$, $M_{\star} = 2.7 \times 10^{7} M_{\sun}$ and $f_g =
0.75$, red) and model C1 ($M_{\mathrm{vir}} = 1.4 \times 10^{10} M_{\sun}$, $M_{\star} = 1.1 \times 10^{7} M_{\sun}$,
$f_g = 0.9$, blue). The bottom panel of Fig.~\ref{SFRgsmiso} shows
that although the more massive dwarf galaxy model B has $2.5$ times more mass
in dark matter and stars initially, the lower mass system model C1 actually has
a slighty higher gas mass ($M_{\mathrm{gas}} = 9.9 \times 10^{7}
M_{\sun}$ versus $M_{\mathrm{gas}} = 8.2 \times 10^{7} M_{\sun}$, due
to the higher gas fraction) and a higher star formation rate. After 2
Gyr the stellar masses of the two systems are similar and also the gas
fractions are more comparable than initially.
 
\begin{figure*}[!ht] 
\includegraphics[width=\textwidth]{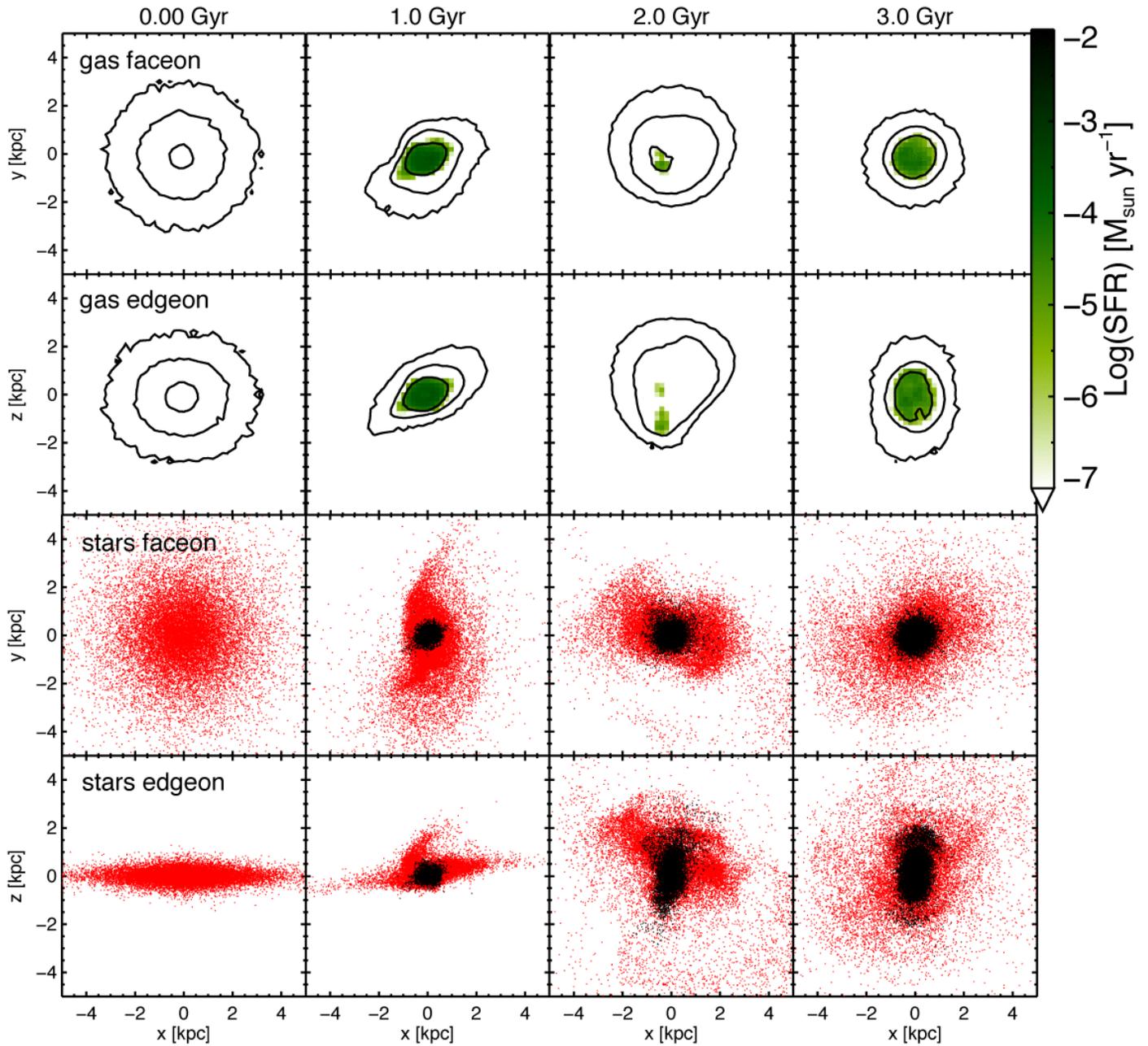}
\caption{\label{D1} Same as Fig.~\ref{merger} for model D1 ($M_{\star}=4.4 \times 10^6 M_{\sun}$ with $R_g=R_d$), merging with a 20\% mass satellite on a 30-degrees inclined, very radial orbit.}
\end{figure*}

Although the simulated systems have slowly declining star formation
rates in general (as a result of the lack of external gas infall), they agree quite well with observations, also in terms of
gas content \citep{Hunteretal2006,Weiszetal2012,Ottetal2012,Warrenetal2012,Huangetal2012,McQuinnetal2015}. However, the lower mass disks tend to be more extended,  slightly thinner,
and have lower surface brightness compared to
observations. We make a more detailed comparison with observations of the isolated and
merger remnant dwarf galaxies in
Sect.~\ref{Comparisonobs}.

\section{Starbursting dwarfs as the result of an interaction}
\label{Results}

In this Section we compare first the behaviour of one of the highest (model A) and one of the lowest (model D1)
mass dwarfs, as they interact with a dark satellite. We describe the general properties and then address in
more detail how the interacting systems vary depending on the
properties of the satellite, its orbit, and the host. In all
cases explored the evolution is quite different from the dwarfs in
isolation, as the majority of the systems experience starbursts of
varying strength.

\subsection{Two examples}
\label{TwoEx}

Figure~\ref{merger} shows a series of snapshots of the evolution of
the system of Fig.~\ref{iso}, model A ($M_{\star} = 1.4 \times 10^8 M_{\sun}$) with $R_g = 2 R_d$, now
merging with a satellite on a co-planar very radial orbit. The stellar
disk thickens, tilts, and develops tidal arms and rings. The gaseous disk 
depicts minor tidal arms and becomes quite asymmetric. Most
importantly gas is driven into the center of the dwarf galaxy due to
tidal torques, which leads to a strong increase of the star formation
rate as we discuss below.

Fig.~\ref{D1} shows the evolution for the lower mass model D1 ($M_{\star}=4.4
\times 10^6 M_{\sun}$) merging with the satellite on a very radial 30
degrees inclined orbit. Similar to what was found in \citet{SH15}, for
this smaller mass object the effect of the satellite on the stellar
disk is much stronger. The stellar morphology of
the remnant is spheroidal, for both the old and the newly formed
stellar populations (see the rightmost panels of Fig.~\ref{D1}). The gas
disk is severely disturbed, although no strong tidal tails
form. As can
be seen in the third panel of the top two rows, the presence of the
satellite can cause off-center starbursts. 

\begin{figure} 
\includegraphics[width=0.5\textwidth]{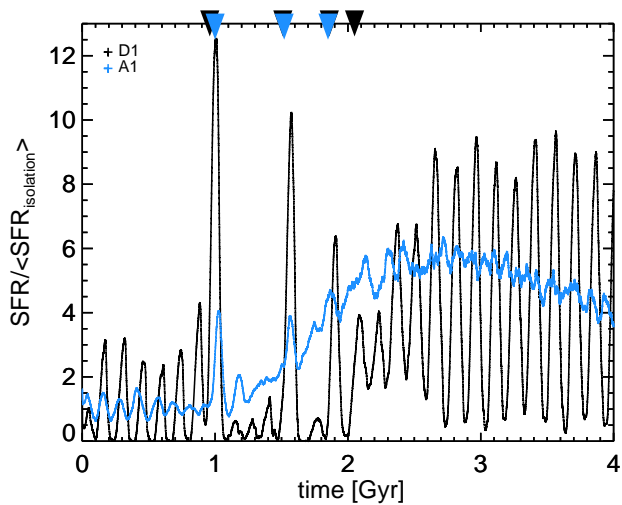}
\caption{\label{SFR2examples} Star formation rates with respect to the average of the SFR in isolation for the first Gyr, for the systems A1 (see Fig.~\ref{merger}) and D1 (see Fig.~\ref{D1}) experiencing a 20\% mass ratio merger.A number of pericentric passages of the satellite during the different simulations are indicated by the arrows at the top of the figure.}
\end{figure}

Fig.~\ref{SFR2examples} shows clearly the effect of the merger on the
SFR for these two systems.  Both for model A (blue) and D1 (black) the
star formation rates are increased by factors $\sim$ 3 up to 12. This happens not only during the merger process itself with
sharp strong peaks at or just after pericenter passages, but more strinkingly for
a rather extended period. The relative amplitude of the increase in the
SFR is similar for both systems, although in an absolute sense, the
more massive object naturally has a higher SFR, reaching values of
$0.04 M_{\sun}$ yr$^{-1}$. Note as well from Figs.~\ref{merger} and
\ref{D1} that in both cases the gas disk at the end of the
simulation is much more compact than initially.

\begin{figure}
\includegraphics[width=0.5\textwidth]{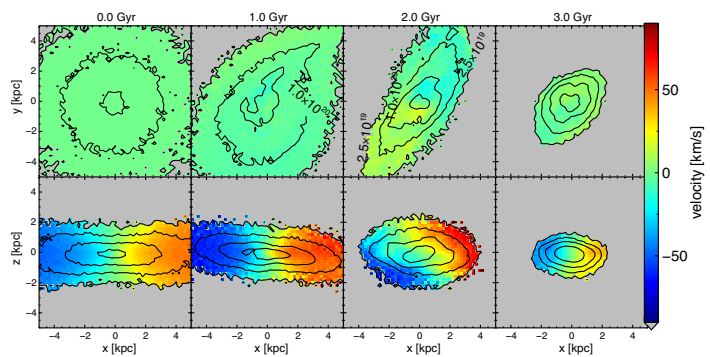}
\caption{\label{vel-merger} Face-on (top) and (bottom) edge-on gas
  contours for model A with $R_g = 2 R_d$ merging with a 20\% mass
  satellite on a radial, co-planar orbit depicted in
  Fig.~\ref{merger}, initially (far left) and after 1 (middle left), 2
  Gyr (middel right) and 3 Gyr (far right). The plane of the disk in
  determined by the angular momentum of the inner 50\% of the initial
  stellar particles, so the rotation in the gas can be in a different
  plane.}
\end{figure}

\begin{figure} 
\includegraphics[width=0.5\textwidth]{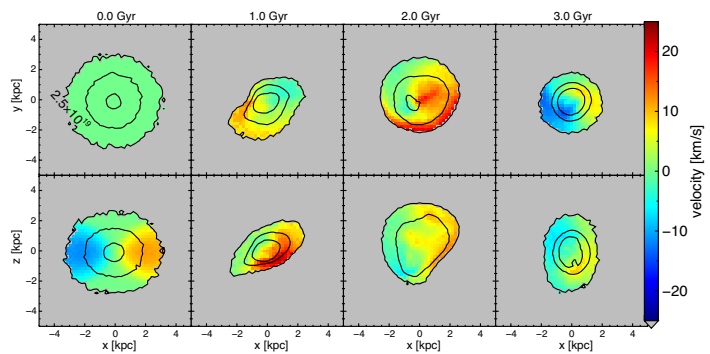}
\caption{\label{vel-D1} Same as Fig.~\ref{vel-merger} now for the D1 system depicted merging with a satellite as shown in Fig.~\ref{D1}.}
\end{figure}

In Figures~\ref{vel-merger} and~\ref{vel-D1} we plot the evolution in
the rotational velocity for the two systems shown in Figs.~\ref{merger}
and~\ref{D1}, respectively. The rotational motion of the gas disk of
model A seems to increase first (see the left middle panel of
Fig.~\ref{vel-merger}) and then strongly decreases as the gas disk
shrinks until the significantly smaller maximum circular velocity $\sim 30$ km/s is reached. Nonetheless, the velocity field
generally remains rather robust during the merger.  For the lower
mass system, which depicts a smaller amplitude of rotation initially,
the velocity field is much less conspicuous and ordered during and after 
the merger. Note that at the end, the maximum rotation signal is found for
what we have defined as the ``face-on'' view of the system, although
this characterization is debatable given the spheroidal shape of the
remnant.


\subsection{Variation in properties of the satellite}

For a significant effect, the satellite must reach the stellar and gas disks of the dwarf galaxy within a short timescale to be apparent in our simulations. Satellites on orbits that are close to circular will take a longer time to sink to the center. This is longer than the few Gyr run-time of our non-cosmological simulations to limit environmental and cosmological effects, such as the lack of cosmic gas inflow. Moreover, the strength of the perturbation depends on the average density ratios of the satellite to the host. To explore the dependencies of the mergers on the satellite's properties we consider a number of different orbits, and satellites with different concentrations (and central densities), for three different mass ratios: $1:5$ (our default), $1:10$, and $1:20$. 

\subsubsection{The satellite orbit}
\label{orbit}

\begin{figure} 
\includegraphics[width=0.45\textwidth]{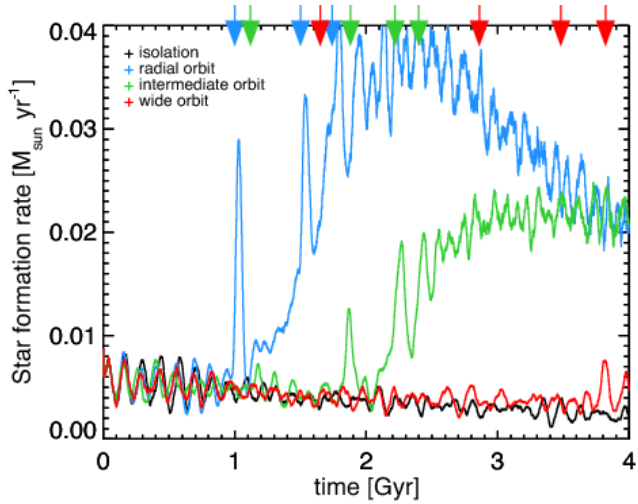}
\caption{\label{SFRH2orbits} Star formation rates for model A with $R_g = 2 R_d$: 
  in isolation (black) and experiencing a 1:5 merger with a high concentration, $c=25$, satellite on a planar orbit: radial with first pericenter within the 
  stellar disk (blue), less radial with first pericenter just outside the stellar disk (light green), and even less radial with $r_{\mathrm{peri}} \approx 20\ \mathrm{kpc}$ (red). A number
  pericentric passages of the satellite during the different simulations are indicated by the arrows at the top of the figure.} 
\end{figure}

Figure~\ref{SFRH2orbits} compares the star formation rates for the
dwarf galaxy model A experiencing a 1:5 merger with the same (high concentration)
satellite on three different orbits. For less radial orbits the disk
forms large tidal spiral arms (compare for example Figs.~\ref{merger-radial-highC}  and \ref{merger-lessradial-highC} in the Appendix) and the subsequent increase in star formation only starts when
the satellites comes within $\sim 3\ \mathrm{kpc}$ of the center. For the most extreme
example shown here (the red line in Fig.~\ref{SFRH2orbits}) this
happens only at $3.5$~--~$4$ Gyr, i.e. at the end of the simulation run.

\begin{figure} 
\includegraphics[width=0.45\textwidth]{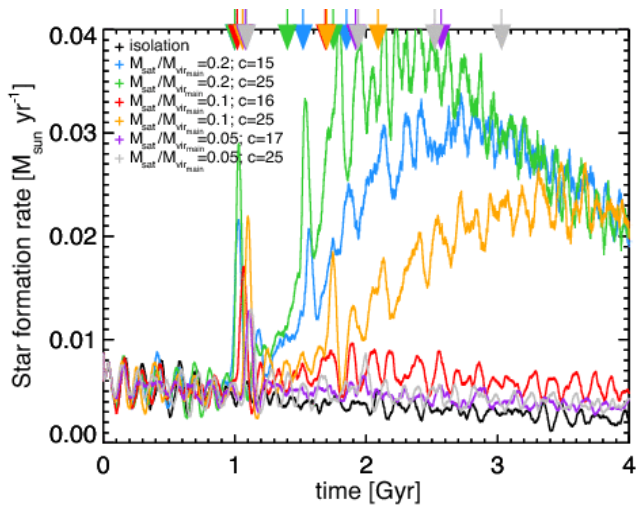}
\caption{\label{SFRH2sats} Star formation rates for model A with $R_g = R_d$: in isolation (black), and with a satellite on a planar radial orbit with $M_{\mathrm{sat}}/M_{\mathrm{vir_{\rm main}}}=0.2$ ($c_{\mathrm{sat}} = 15$ blue; $c_{\mathrm{sat}} = 25$ green), $M_{\mathrm{sat}}/M_{\mathrm{vir_{\rm main}}}=0.1$ ($c_{\mathrm{sat}} = 16$ red; $c_{\mathrm{sat}} = 25$ orange), and $M_{\mathrm{sat}}/M_{\mathrm{vir_{\rm main}}}=0.05$ ($c_{\mathrm{sat}} = 17$ purple; $c_{\mathrm{sat}} = 25$ gray). A number pericentric passages of the satellite during the different simulations are indicated by the arrows at the top of the figure.}
\end{figure}

For more circular orbits the spiral arms that are generated in the gas (and stellar) disk are more pronounced. Nevertheless, for the radial orbit the gas is much more concentrated after the merger, which suggests that more gas is funnelled to the center due to tidal torques. 

We also experimented with different inclinations for the satellite orbit with respect to the plane of the disk (see Figs.~\ref{SFR-A-incl} and \ref{SFR-C1-incl} for examples). In general we find that an inclination of 30 degrees has a larger effect on the morphology and kinematics of the stellar disk \citep[see also our collissionless simulations described in][]{SH15} but drives less gas to the center and produces a smaller increase in star formation rates compared to a co-planar orbit. 

\subsubsection{Satellite mass and concentration}
\label{satmass}

For our default setup the satellite mass is 20\% of the virial mass of
the host dwarf galaxy. This choice is motivated by the fact that
interactions with such objects can be devastating and may not be so
rare.  \citet{Helmietal2012} have estimated that dwarf galaxies
experience on average 1.5 such encounters in a Hubble time, but
encounters with smaller mass objects are certainly more common in the
$\Lambda$CDM cosmological model.

\begin{figure} 
\includegraphics[width=0.5\textwidth]{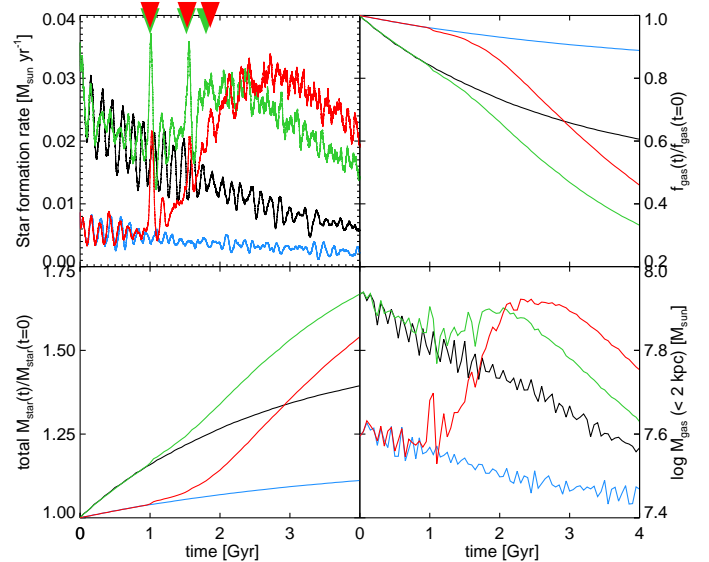}
\caption{\label{SFRgsmmergerH12} Star formation rates, relative gas
  fractions, relative stellar masses, and gas masses in the central
  parts for model A with $R_g = R_d$ and $R_g = 2 R_d$ in isolation
  (black and blue, respectively) and during the merger (green and red,
  respectively). The arrows at the top of the top-left figure indicate
  a few pericentric passages of the satellite during the simulations
  shown.}
\end{figure}

\begin{figure}[!h] 
\includegraphics[width=0.5\textwidth]{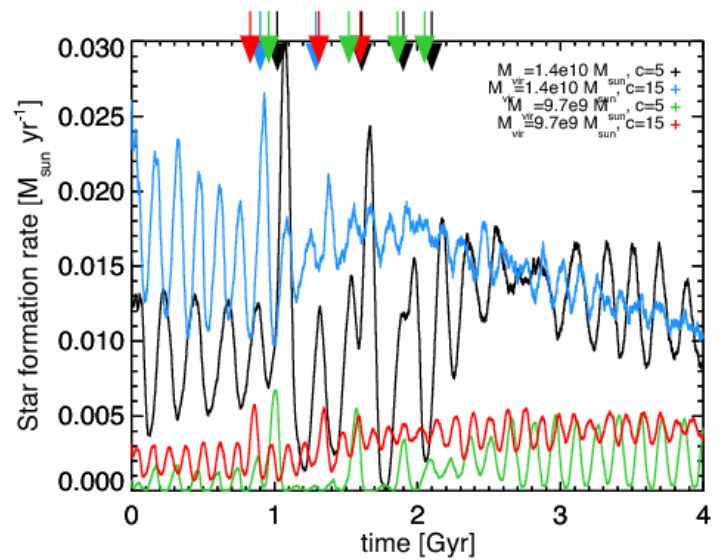}
\caption{\label{SFRmergerfg0.9} Star formation rates for models C and
  D with $R_g = R_d$, $f_{\mathrm{gas}} = 0.9$, $z_0 = 0.3 R_d$, and a
  dark matter halo concentration of either $c=5$ (C3 and D2) or $c=15$
  (C1 and D1), during the 1:5 merger with a satellite on a prograde
  orbit with an inclination with the plane of the disk of 0 (for models C) or 30
  degrees (for models D). A number pericentric passages of the satellite during the
  different simulations are indicated by the arrows at the top of the
  figure.}
\end{figure}

We explore the effect of 1:10 and 1:20 mergers for the most massive dwarf galaxy (model A) with $R_g=2R_d$, and place the satellites on a radial, planar orbit. We consider satellites following the mass--concentration relation from \citet{MunozCuartasetal2011} and also having a higher concentration of $c_{\mathrm{sat}}=25$. 

\begin{figure*}[!ht] 
\includegraphics[width=\textwidth]{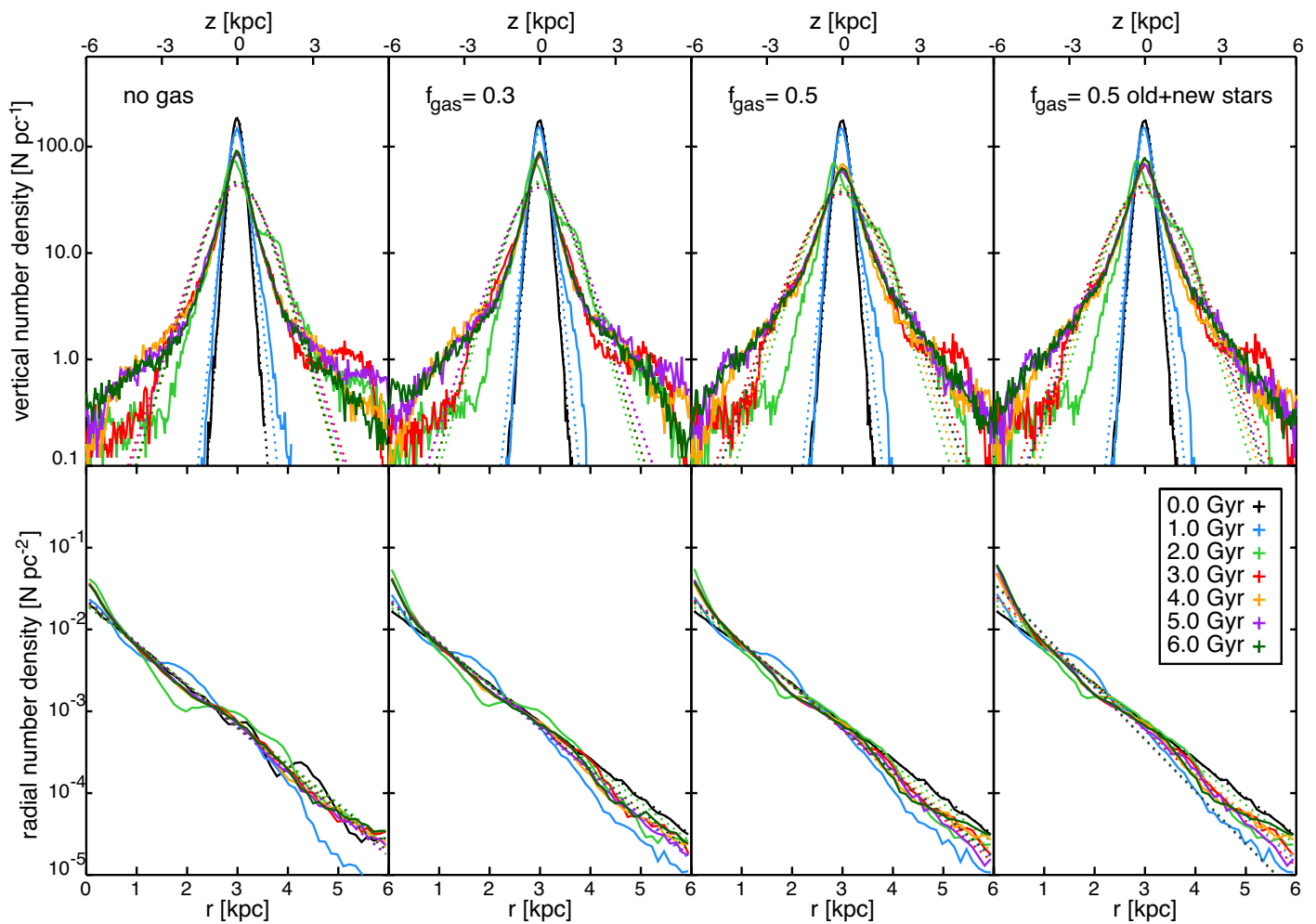}
\caption{\label{profilescollgas} Vertical (top panels) and radial (bottom panels) density profiles (solid lines) during the merger simulations with intervals of 1 Gyr for the \emph{FNX}-like system: fully collissionless (left), with $f_{\mathrm{gas}} = 0.3$ and $f_{\mathrm{gas}} = 0.5$ (middle), and with $f_{\mathrm{gas}} = 0.5$  including the newly formed stars (right). The dotted lines show the maximum likelyhood fitted profiles:  isothermal (sech$^2$)  and exponential in the top and bottom panels, respectively.} 
\end{figure*}

The resulting SFR are shown in Fig.~\ref{SFRH2sats}. In all cases the amplitude of the starburst depends most strongly on the mass and secondly on the concentration of the satellite, although the latter dependance is weaker the lower the mass of the satellite. Note also that the onset of the major starburst is later for smaller mass satellites. This is because the satellite sinks in more slowly, i.e. the pericenter passages where star formation is triggered occur later. 

For all satellites, higher concentration leads to slower mass loss,
hence to more damage to the host. This results in a larger
increase in its SFR and to a stronger morphological disturbance of  stellar and gaseous disks, with
the final gas distribution being more centrally concentrated. While gas gets blown out
of the disk in all cases, in the interaction with the highest mass and $c_{\mathrm{sat}}=25$ satellite the
gas reaches $\sim 4 r_{\mathrm{vir}}$, about twice as far as for the
other cases.

In summary, in all cases (although very minor for $M_{\mathrm{sat}}/M_{\mathrm{vir_{\rm main}}} = 0.05$), there is a first strong peak of star formation at the first pericentric passage of the satellite, and a more extended in time starburst, also driven by the merger. Secondary peaks associated to subsequent pericentric passages are also present but are generally less conspicuous.

\subsection{Influence of properties of the host dwarf galaxy}

The structure of the dwarf galaxy itself can significantly alter the effects of the minor merger on the gas and star formation. We vary the halo mass and concentration, the gas fraction and extend of the gas disk, and the scale length and scale height of the stellar disk. Of these, varying stellar disk parameters causes only very minor differences on the gas and star formation.

\subsubsection{Gas distribution}
\label{fgRgRd}

We explore now the evolution of a disky dwarf galaxy when the gas disk has different extent than the stellar disk, since this characteristic is often seen in large spiral galaxies. A comparison between the merger properties for model A with $R_g = 2 R_d$ and with $R_g = R_d$ can be seen in Fig.~\ref{SFRgsmmergerH12}. 

Due to the lower initial star formation rate for the more extended (and hence lower surface density) gas disk, this system has a larger gas reservoir at the time of the merger (see top right panel). The bottom right panel of Fig.~\ref{SFRgsmmergerH12} shows that the increase in gas mass near the centre is much higher for the initially more extended disk during the merger (red curve) than for the less extended gas disk (green). However, the final increase in stellar mass is still higher for the system with $R_g =R_d$, mostly due to its initially larger star formation rate. 

In general more extended disks have lower star formation rates both in isolation and during the merger. This is because the same amount of gas is distributed over a larger area and therefore the amount above the star formation threshold is much lower.

\subsubsection{Host's dark matter halo mass and concentration}
\label{halomassc}

Figure~\ref{SFRmergerfg0.9} shows the star formation rates for the
dwarf galaxies models C1 and C3, and D1 and D2 which have different
dark matter halo concentration and mass. Even though the difference in
halo mass between models C and D is only a factor $1.4$, due to the steepness of the halo
mass~--~stellar mass abundance matching relations
\citep{Behroozietal2013,Mosteretal2013,Garrison-Kimmeletal2014,Sawalaetal2015}
the difference in stellar mass and gas mass is a factor $2.5$. For
these runs the stellar disk is thick initially ($z_0 = 0.3 R_d$), $R_g
= R_d$, $f_{\mathrm{gas}} =
0.9$, and the satellite has a mass of 20\% of the host halo on
radial orbit with an inclination of 0 or 30 degrees.

As noted before, higher (central) mass implies initially higher SFR,
and thus also during and after the merger (compare black and green, and blue and red curves). A less concentrated host
halo initially has a puffier gas disk and lower SFR, but during a
merger the enhancement in the SFR is 
larger, as depicted by the black curve in Fig.~\ref{SFRmergerfg0.9}
for model C3. Therefore we see that the susceptibility of the system to a merger depends both on virial mass of the host and on its concentration.

\section{Comparison with collissionless runs}
\label{Comparisoncoll}

In this section we compare the dwarf galaxies models C4 and C5
(\emph{disk3-gas}) and E1 and E2 (\emph{FNX-gas}) with the results from
the ``equivalent'' collissionless simulations in \citet{SH15}. Our
focus is on how the presence of the gas changes the characteristics of
the merger remnants.

The morphological and kinematical changes to the stellar component of
both dwarf systems are very similar to the
fully collissionless runs. We exemplify this in Fig.~\ref{fnx-fg0.5} of the Appendix, and more quantitatively in Figure~\ref{profilescollgas}. This figure shows the
vertical and radial density profiles of the stellar disks of the
initially thick \emph{FNX-analog} system from \citet{SH15} (left
panels) and for the counterparts with $f_g = 0.3$ and 
$f_g = 0.5$ (central panels). In all three
cases the strongest changes occur during the first pericenter passages
of the satellite, within 2 Gyr of the start of the simulations. The
differences in the structural evolution of the stellar disks are
surprisingly small. The major difference is that for the fully collisionless system the radial
profile of the disk is slightly more unstable and has more
substructure. Even the newly formed stars in the runs
including gas do not significantly change the stellar density profiles
as can be seen in the rightmost panels of Fig.~\ref{profilescollgas}, but
they give rise to a small central bulge (compare the bottom panels). For the more massive \emph{disk3} 
(models C4 and C5) the inclusion of gas also has a small effect, and leads to a slightly thicker stellar disk in
the post-merger phase, more so for higher gas fractions.

The effect of the merger on the kinematics of the stars do not differ
much when gas is included in the disk. For example, for the \emph{FNX}-like system
in both cases the rotation decreases over the whole disk while the
circular velocity increases due to the accretion of the dark matter
satellite. All velocity dispersions increase, and even increase slighly
more in the presence of a gas disk. This might be due to the fact that
the gas disk is not very thin due to the low mass of the dwarf galaxy
halos and therefore exerts less of a pull toward the disk midplane
compared to a purely stellar disk.

\begin{figure} 
\includegraphics[width=0.5\textwidth]{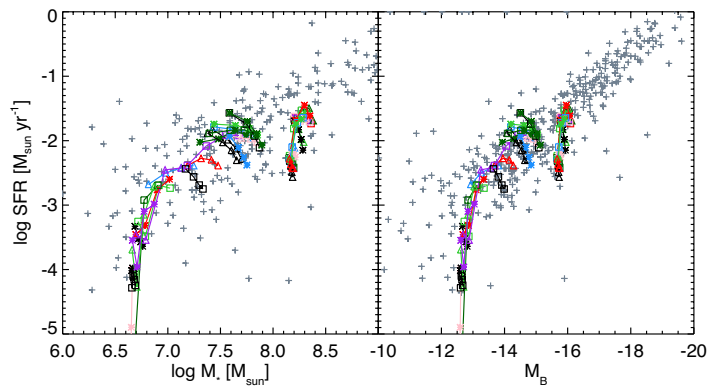}
\caption{\label{Weisz} Star formation rates versus stellar masses and
  B-magnitude in intervals of 1 Gyr in comparison to the dwarf
  galaxies from \citet{Weiszetal2012}, \citet{Huangetal2012} and
  \citet{McQuinnetal2015} for the left panels, and in addition for the
  right panel from \citet{Ottetal2012}, \citet{Warrenetal2012}, and
  \citet{Hunteretal2006}. Note that the stellar mass grows during the
  simulations.}
\end{figure}

\begin{figure}[!t] 
\includegraphics[width=0.5\textwidth]{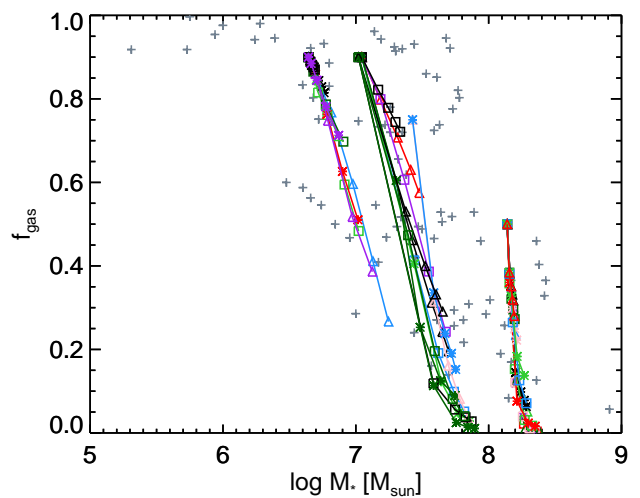}
\caption{\label{fgasmstarsfr} Gas fractions versus stellar masses in intervals of 1 Gyr in comparison to the dwarf galaxies from \citet{Huangetal2012}, and  \citet{McQuinnetal2015}. Since the stellar mass grows during the simulations and that we do not include fresh gas inflow, the gas fraction of our simulated systems necessarily decreases.}
\end{figure}

Therefore, the presence of gas in the disk has no significant
influence on the effects of minor mergers on dwarf galaxies. This is
in striking contrast to what happens in similar simulations for higher
(Milky Way) mass galaxies (see discussion in Sect. \ref{Discussion}).

\section{Comparison to observations}
\label{Comparisonobs}

We now compare the properties of our simulated dwarf galaxies, in isolation and during the merger, with the observational samples described in Table~\ref{obs}. This comparison includes systems from HI-selected, as well as mass-selected samples and general compilations.

   The left panel of Fig.~\ref{Weisz} shows the SFR versus stellar
   mass $M_*$ for the systems discussed in this paper compared to
   observational values reported in \citet{Weiszetal2012},
     \citet{Huangetal2012}, and \citet{McQuinnetal2015}. For each simulation the
   evolution of the SFR and $M_*$ are plotted at four epochs in steps of 1 Gyr, from 1 Gyr after the start of the
   simulation. The black symbols correspond to the runs
   in isolation while in color we show the mergers.  Overall the star
   formation rates match very well those from observations for their
   stellar masses. 

   To estimate the luminosities and surface brightness of our
   simulated galaxies we need to assume mass-to-light ratios for the
   stellar particles. One possibility for the stellar particles formed
   during the simulation would be to use stellar population models as
   we know their ages. For the stellar particles initially present in
   the simulation however the age distribution would be
   arbitrary. Therefore, we instead fitted a linear relation to the
  $\mathrm{log}\ M_{\star}$--B magnitude distribution for the galaxies in the
   observations of \citet{Weiszetal2012} and
   \citet{McQuinnetal2015} and apply it to the simulated dwarf galaxies.  The resulting 
   estimated values are roughly consistent with an average mass-to-light ratio of  
   $M_{\star}/L_{B,\star} \sim 0.5$, which is quite reasonable as
   the light is very strongly dominated by young stars. The right panel of
   Fig.~\ref{Weisz} shows SFR against the B-magnitude for our systems and for a larger sample of
   observations. Good agreement is also found in this case. 

 \begin{table*}[\textwidth]
      \caption{\label{obs} Observational samples used}
         $$ 
         \begin{tabular}{llr}
            \hline

            \noalign{\smallskip}
               Reference & Sample & Properties used \\    
            \noalign{\smallskip}
            \hline
            \hline
            \noalign{\smallskip}
            \noalign{\smallskip}
\citet{Hunteretal2004} and & 94 Im, 24 BCD, and 18 Sm & $M_{\mathrm{HI}}$, $M_V$, $B-V$, $R_d$, SFR, $\mu_0$ \\
\citet{Hunteretal2006} & & \\
\citet{Weiszetal2012} & 185 galaxies from  & $M_{\star}$, $M_B$, SFR \\
 & the \textit{Spitzer} LVL survey \citep[e.g.,][]{Daleetal2009}) & \\
 & and 11HUGS \citep{Kennicuttetal2008, Leeetal2011} & \\
\citet{Huangetal2012} & 229 low HI mass galaxies from & $M_{\star}$, $M_{\mathrm{HI}}$, SFR \\
 & the ALFALFA survey \citep{Giovanellietal2005} & \\
\citet{Ottetal2012} & VLA-ANGST survey: 35 galaxies from & $M_B$, SFR \\
 & the ANGST survey \citep{Dalcantonetal2009} & \\
\citet{Warrenetal2012} & 31 nearby low-mass galaxies taken from & $M_B$, SFR \\
 & THINGS \citep{Walteretal2008} and & \\
 & the VLA-ANGST survey \citep{Ottetal2012} & \\
\citet{McQuinnetal2015} & 12 galaxies from the SHIELD survey \citep{Cannonetal2011} & $M_{\mathrm{HI}}$, $M_{\star}$, SFR, $M_B$ \\
\citet{Karachentsevetal2004} & all-sky catalog of basic optical and HI properties of & $v_{\mathrm{rot}}$, $b/a$ \\
 & (451) neighboring galaxies with $D < 10\ \mathrm{Mpc}$  & \\
 & or radial velocity $V_{\mathrm{LG}} < 550\ \mathrm{km/s}$ & \\
            \noalign{\smallskip}
            \hline
            \hline
         \end{tabular}
     $$ 
   \end{table*}

\begin{figure} 
\includegraphics[width=0.5\textwidth]{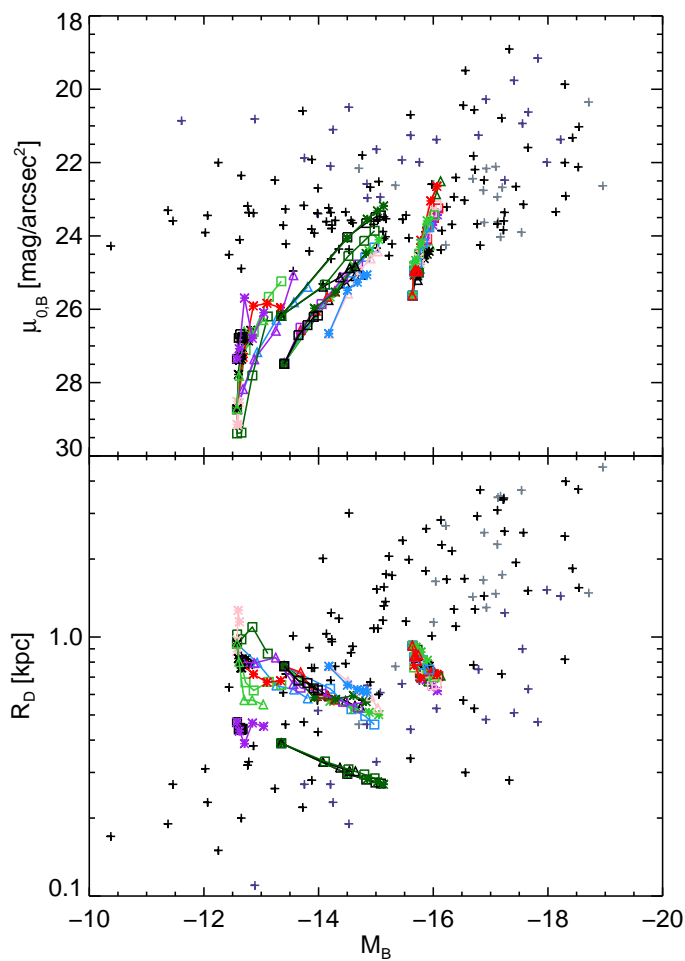}
\caption{\label{Hunterdisks} Central surface brightness (top) and disk scale lengths (bottom) versus absolute magnitudes in intervals of 1 Gyr compared to irregular dwarf galaxies (black), blue compact dwarfs (dark blue), and Magellanic spirals (gray) from \citet{Hunteretal2006}. Note that the stellar mass grows (mostly in the center) and so the magnitude and disk scale length decreases and the central surface brightness increases during the simulations.}
\end{figure}

\begin{figure} 
\includegraphics[width=0.5\textwidth]{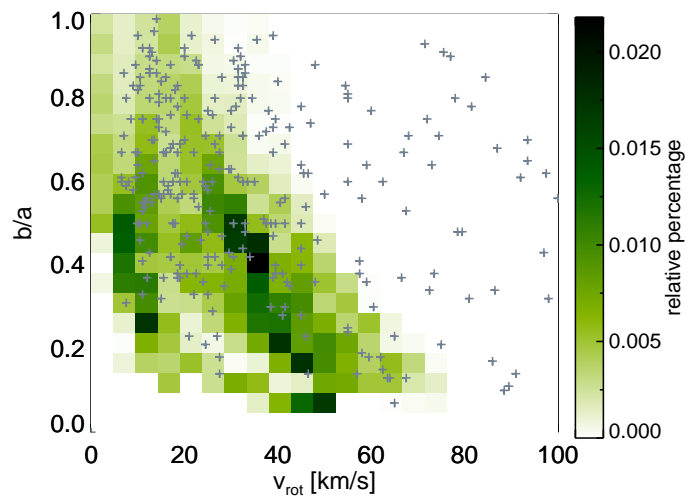}
\caption{\label{axisratiosstat} Distribution of ``projected'' axis ratios and rotational velocity for 100 random inclinations at intervals separated by 1 Gyr during each of the simulation runs, compared to galaxies with $v_{\mathrm{rot}} = 0.5 \times w_{50} \leq 100\ \mathrm{km/s}$ from the catalog of nearby galaxies \citep{Karachentsevetal2004} (gray). Darker colors indicare a higher density of objects in this plane.}
\end{figure}

   Figure~\ref{fgasmstarsfr} shows that also the initial gas fractions
   and their evolution in time agrees very well with observations. The
   downward trend seen in the simulated systems is due to the fact
   that we do not include fresh gas inflow. However, sampled at an arbitrary point in time, the match is quite remarkable.

We compute radial surface brightness profiles for the face-on disks and fit 
an exponential profile to derive the disk scale
length and half-light radius. This process is similar to that of
\citet{Hunteretal2006} and we compare our disks with their
observational results. We derive central surface brightness using the mass/luminosity within the innermost, $0.1$ kpc, bin in projection. The results are shown in Fig.~\ref{Hunterdisks}. It is clear that
while the higher mass disks match very well with the observations, the
lower mass systems seem to be more extended and fainter, compared to
the observations. There are two main possible reasons for this: low
surface brightness, extended systems are harder to observe. On the
other hand, the method we used to compute the initial size of the
systems and to set up our simulations \citep[using the disk mass and
properties of the halo following][]{MoMaoWhite1998} might break down
for lower mass systems, as they typically are thicker and deviate from being
thin disks. To estimate the effect of this we have also run the two lower
mass systems with initial disk scale lengths half their default values and, not surprisingly, we find that they agree better.

In Fig.~\ref{axisratiosstat} we compare our ``observed'' axis ratios
and rotational velocities to the observed axis ratios of all galaxies
with $w_{50} < 200\ \mathrm{km/s}$ from the catalog of nearby galaxies (450 when the
Milky Way is excluded) by 
\citet{Karachentsevetal2004}.  The intrinsic axis ratios are computed
by fitting an ellipsoid to the stellar particle distribution
\citep[normalized by the ellipsoidal distance of the particles within
this ellipsoid,][]{Allgood2006}. The ellipsoids are centered around
the center of mass defined by the stellar component and the axis ratios
computed for particles within major axis length of $a \sim 1.4\
\mathrm{kpc}$, but we find similar results when considering particles
within $a \sim 4\ \mathrm{kpc}$ and even $a\sim 14\ \mathrm{kpc}$.
These intrinsic minor-to-major axis ratios are in agreement for most
of the merger remnants, but the default initial disks are often
thinner than the observations\footnote{Initially thicker disks ($z_0 = 0.5
R_d$), which might be expected for dwarf galaxies that form in gas
that has cooled less efficient than in larger disks
\citep{Kaufmannetal2007, RobertsonKravtsov2008}, already agree well
with the observational points.}. However, most systems will not be
observed edge-on, and therefore to provide more realistic estimates of
``observed'' axis ratios we put each system at 100 random inclinations and compute
projected axis ratios at five evenly spaced
epochs during the simulation.

For the rotational velocities, we assume that for the observational sample $v_{\mathrm{rot}}
\approx 0.5 \times w_{50}$ and compute the rotational velocities for
our modelled systems as the mean between the minimum and maximum value
of the velocity maps (as in Fig.~\ref{vel-merger}), which give the velocities in the disk
within the column density contour of $2.5 \times 10^{19}\ \mathrm{N}\ \mathrm{cm}^{-2}$ (which is close typically to $\sim 2 R_D$).

Inspection of Figure~\ref{axisratiosstat} shows when including the
(random) inclination angle, the sample of simulated dwarf galaxies
agrees very well with the observed sample. Surprisingly, even the
trend that the lower-mass stellar disks are more extremely perturbed
by the minor merger is apparent.

\textit{Overall, our systems agree well with properties of observed dwarf galaxies.}

\section{Discussion}
\label{Discussion}

In this paper we show the effects of minor mergers with a dark
satellite on gas-rich dwarf galaxies. The isolated systems a carefully
set up and we have explored the effects of varying the subgrid
parameters on the results. Our star formation and feedback
prescriptions lead to reasonable dwarf galaxies when evolved in
isolation, and their properties are fairly stable against varying these
parameters. Only extremely efficient feedback gives a completely
different and unrealistic evolution. We have also carried some of the
merger experiments for different feedback schemes and found
that the effects of the merger on the gas and star formation evolution 
to be very robust against such changes. 

When comparing the fully hydrodynamic models to the collissionless
minor merger simulations of \citet{SH15} we have found that the evolution of the
stellar components of the dwarf galaxies is rather similar. Even though the
gas absorbs and dissipates some of the energy injected via the
merger, the morphological transformations that the stellar disks experience are still very
important. This is in contrast to what has been reported in the
literature for Milky
Way-size systems where the effect is significantly reduced when gas is included \citep[e.g.,][]{Hopkinsetal2009, Mosteretal2010}. This is probably due to the gas being much colder in
larger disk galaxies and therefore having a stronger stabilizing
effect on the stars in the disk.

An important question is how often the process discussed in this paper
would happen for dwarf galaxies in different environments, at
different redshifts and with different masses. As the CDM halo mass
function is almost scale-free \citep[with small differences due to
halo formation times][]{vdBoschetal2005,vdBoschJiang2014}, dwarf
galaxies will have a spectrum of perturbers very similar to that of an
$L_{\star}$ galaxy. \citet{Helmietal2012} estimate that the number of
minor mergers for low galaxy efficiency systems
($M_d/(M_{\mathrm{vir}} \times f_{\mathrm{bar}}) = 5\%$) where the
satellite is at least as massive as the disk at pericenter, with the
pericenter within 30\% of the virial radius, is $\sim 1.5$ within a
Hubble time. The simulations discussed in this paper were for systems
with even smaller galaxy efficiencies experiencing 1:5 and 1:10
mergers with dark satellites, albeit on very radial orbits. This means
that almost every dwarf galaxy should have experienced a minor merger
with major effects during its lifetime. Predictions and their
dependencies on environment, redshift and dwarf galaxy mass will be presented in a
forthcoming paper (Starkenburg et al. in prep.).

Several definitions of a starburst exist in the literature
\citep[e.g.,][]{KnapenJames2009,McQuinnetal2010a,Bergvalletal2015}. The
birthrate parameter, $b$=SFR/$\langle$SFR$\rangle$ is often used, and
compares the current SFR or the peak of the burst to the average SFR
over the past Gyr or even the lifetime of the system.  For the more
massive systems in our simulations the SFR in isolation is very similar
to their initial stellar mass divided by a Hubble time, and the
birthrate goes up to $b\sim3$, or even $b\sim10$, during the merger, depending on the configuration. The lower mass
systems have typically more bursty star formation histories even in
isolation.  We may characterize $b \sim 1 $
between these small bursts, $b \sim $a few during bursts in isolation,
but during the merger we find $b > 10$.  Therefore the
increase in star formation rates the simulated systems experience can be
qualified as starburst events. Also when computing the gas consumption
timescale, $\tau_{\mathrm{gas}} = M_{\mathrm{gas}}$/SFR, there is agreement with the
literature. For the star formation rates in isolation, the gas
consumption timescales are long
$\tau_{\mathrm{gas}}^{\mathrm{isolation}} > 10\ \mathrm{Gyr}$, while
these drop significantly by factors of a few during the merger events.

Although observations suggest that only a small fraction of dwarf
galaxies are currently experiencing a starburst \citep[$\sim 6\%$ according
to][]{Leeetal2009a}, this might be just a fraction of the total
number of dwarf galaxies that experienced starbursts in their lifetimes
\citep{Leeetal2009a}.

\section{Conclusions}
\label{Conclusions}

We performed a suite of controlled, minor merger simulations
between carefully set-up gaseous dwarf galaxies and their (dark)
satellites. These interactions can give rise to a strong increase in
the star formation rates in the dwarf galaxies. The increase is in the
form of large sharp bursts during pericenter passages, as well as of
extended boosts due to tidal torques funnelling gas toward the
center. The gas and stellar disks show severely disturbed morphologies
in most cases, especially for lower mass hosts experiencing a 1:5
merger. The gas disks can develop grand tidal tails and their remnants
depict a much more concentrated final distribution in some cases. For
the lowest mass systems explored ($M_* \lesssim 1.1 \times 10^7
M_{\sun}$), the merger can completely destroy the stellar disk. These
objects become spheroidal-like and have bursty star formation in their
center. In contrast to simulations of Milky Way-like systems, the
presence of gas in the disk of the dwarf galaxies does not diminish
the effect of the merger on the stellar component. Our simulations
that include gas show that the strong heating and evolution of the
stellar disk is almost completely identical to the collisionless case.

We have explored the dependence of the mergers and their remnants on
the host and satellite masses, dark matter halo concentrations,
satellite orbits, gas fractions, and structure of the stellar and gas
disks. To have a significant impact, a dark satellite must have at
least 10\% of the mass of the host. More generally, the strength of
the merger's effects decreases with lower satellite masses, but this
depends on satellite halo concentration: a very dense low-mass dark
satellite can survive longer than one of lower concentration, and
therefore have stronger effects. Orbits in the plane of the disk cause
stronger starbursts, while inclined orbits perturb
the stellar components more effectively. Satellites on radial orbits cause stronger
starbursts than those on more circular ones.

When the hosts have lower concentrations, the merger induces stronger
morphological changes \citep[as in][]{SH15} but in general also 
lower star formation rates. This is because SFR correlates with gas
density, which depends in turn on the total mass distribution in the
region probed by the gas disk. Also the gas fraction
and distribution, as well as stellar disk masses and distributions, determine the amount of gas that has high enough densities for star
formation, thereby directly affecting the
amplitude of the starburst that a dwarf experiences during a merger.

Both our initial systems and their remnants compare well with
the observational properties of a large selection of irregular dwarf
galaxies and blue compact dwarfs. Even systems that are strongly
perturbed as a result of a merger with a dark satellite fall
within the scatter seen in the observations. This implies that such
events might well be happening but may not be fully evident.  We have
yet to identify the ``smoking gun'' of the dark merger
scenario. However, this also shows that the effects of interactions
with dark satellites, which are naturally expected within a CDM cosmology, are likely to
play a role in the diversity of the dwarf galaxy population.

\begin{acknowledgements}
We are grateful to Claudio Dalla Vecchia, Joop Schaye, Carlos Vera-Ciro, Alvaro Villalobos and Volker Springel for providing code. AH acknowledges financial support from the European Research Council under ERC-StG grant GALACTICA-240271 and the Netherlands Research Organisation NWO for a Vici grant.
\end{acknowledgements}

\bibliographystyle{aa} 
\bibliography{MergersGas} 

\begin{appendix}
\section{Additional figures}
\begin{figure*} 
\includegraphics[width=0.9\textwidth]{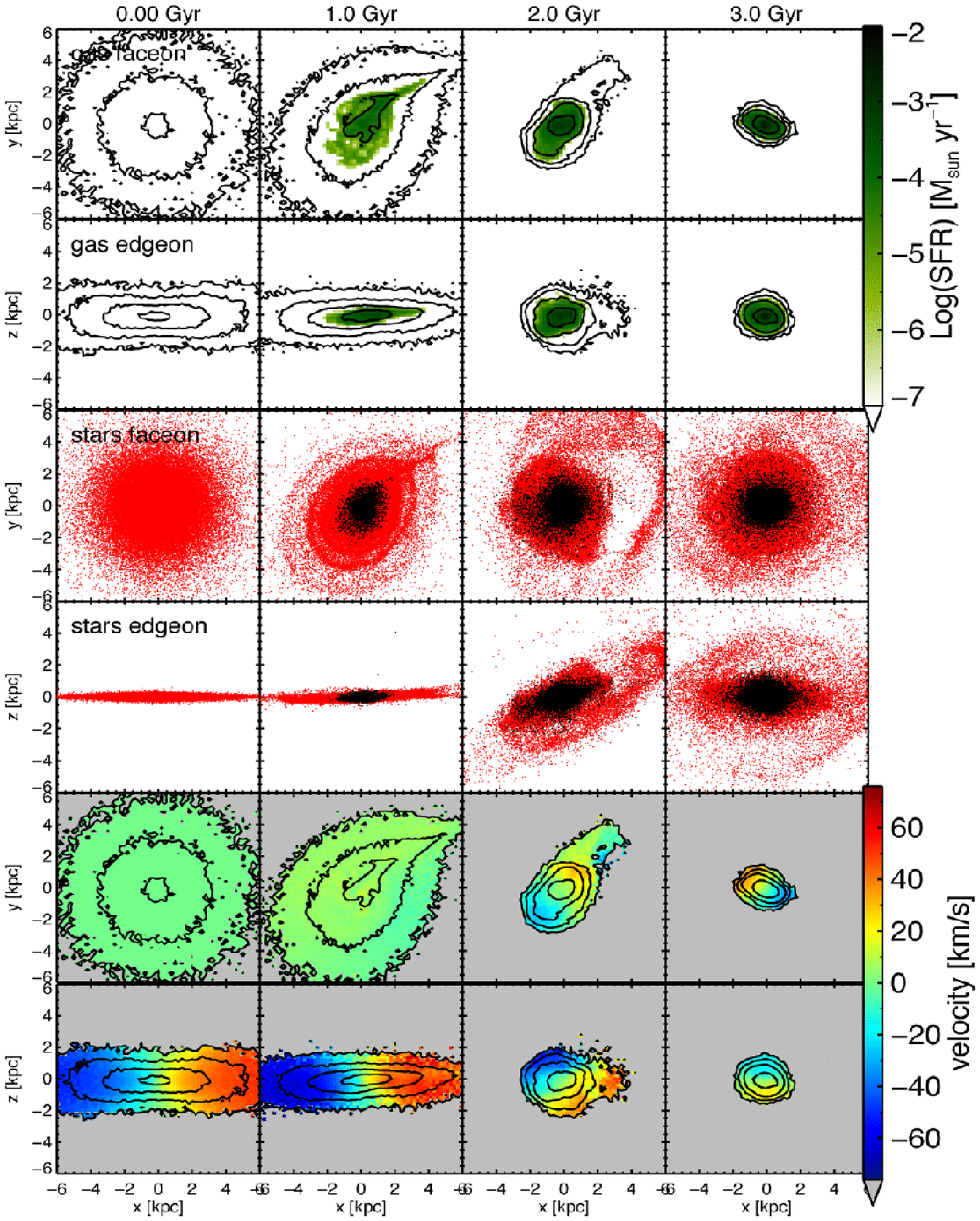}
\caption{\label{merger-radial-highC} Evolution of model A with $R_g= 2 R_d$, merging with the $c=25$ satellite on a planar, very radial orbit, at different times during the merger.  The top rows show the face-on and edge-on, respectively, view of the gas in the disk (contours at $0.25$, $1$, $4$, and $16 \times 10^{20}\ \mathrm{N}\ \mathrm{cm}^{-2}$) with the gas that is currently forming stars highlighted in green (see colorbar for relative values). The third and fourth rows from the top show the old stellar component in red, and newly formed star particles in black. The bottom panels show the gas contours with the gas velocity perpendicular to the contour plane.}
\end{figure*}
\begin{figure*} 
\includegraphics[width=0.95\textwidth]{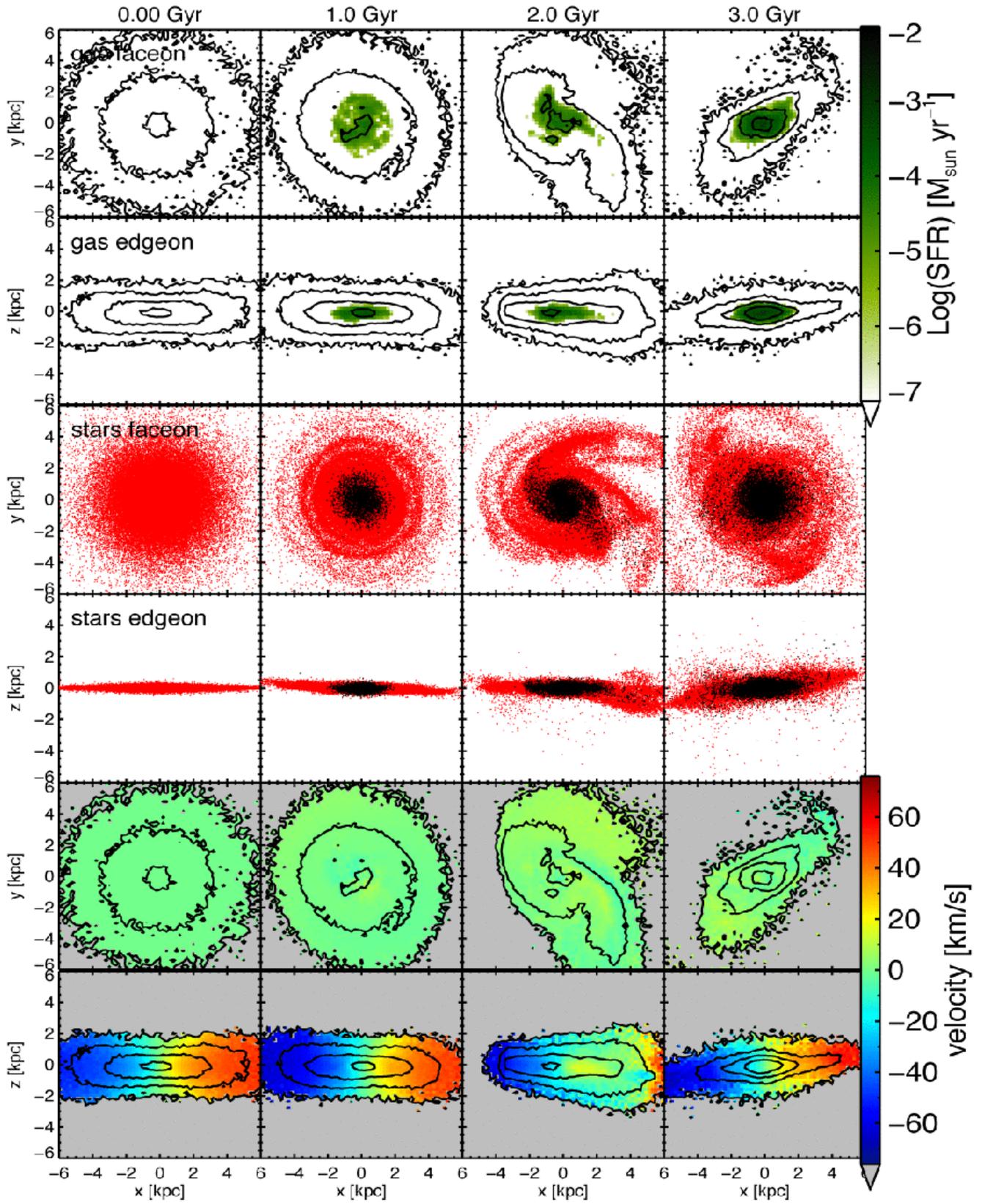}
\caption{\label{merger-lessradial-highC} Same as Fig.~\ref{merger-radial-highC} for model A with $R_g= 2 R_d$, with the $c=25$ satellite on a planar, less radial orbit.}
\end{figure*}

\clearpage
\begin{figure}
\includegraphics[width=0.5\textwidth]{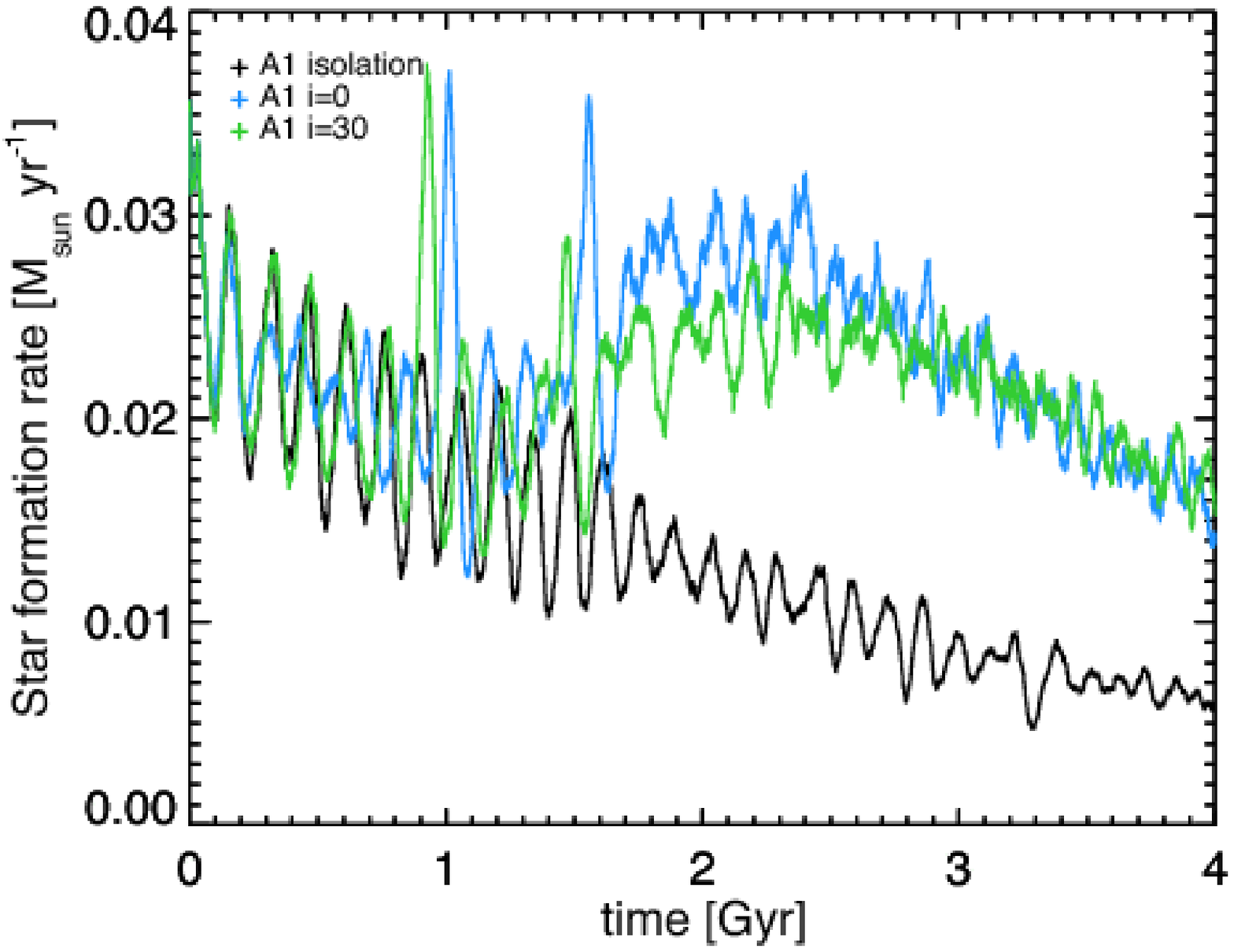}
\caption{\label{SFR-A-incl} Star formation rates for model A with $R_g = R_d$ during the 1:5 merger with a satellite on a prograde, very radial orbit with an inclination with respect to the plane of the disk of 0 or 30 degrees. }
\end{figure}
\begin{figure}
\includegraphics[width=0.5\textwidth]{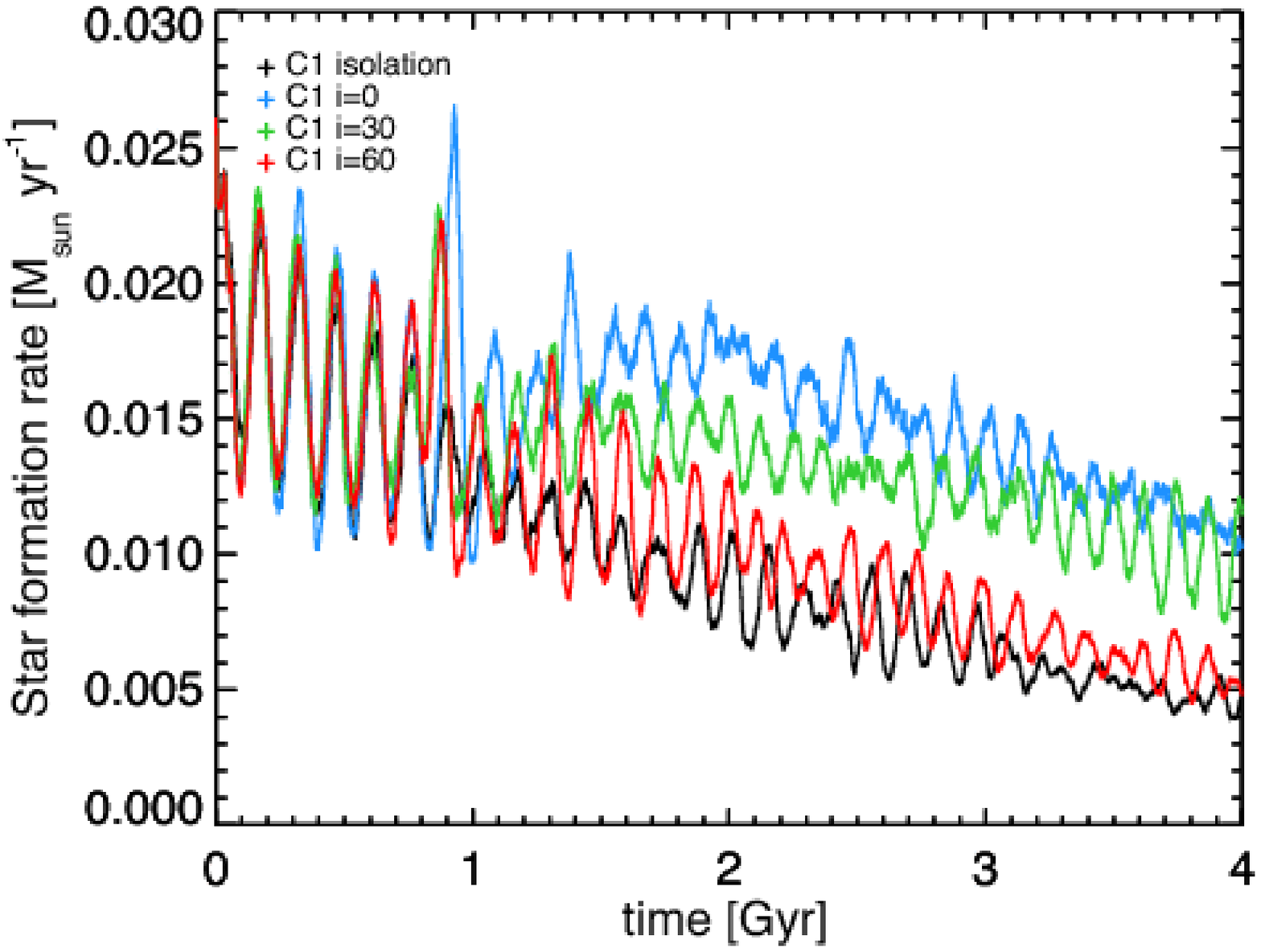}
\caption{\label{SFR-C1-incl} Star formation rates for model C1 with $R_g = R_d$ during the 1:5 merger with a satellite on a prograde, very radial orbit with an inclination with respect to the plane of the disk of 0, 30, or 60 degrees. }
\end{figure}

\begin{figure*}
\includegraphics[width=0.95\textwidth]{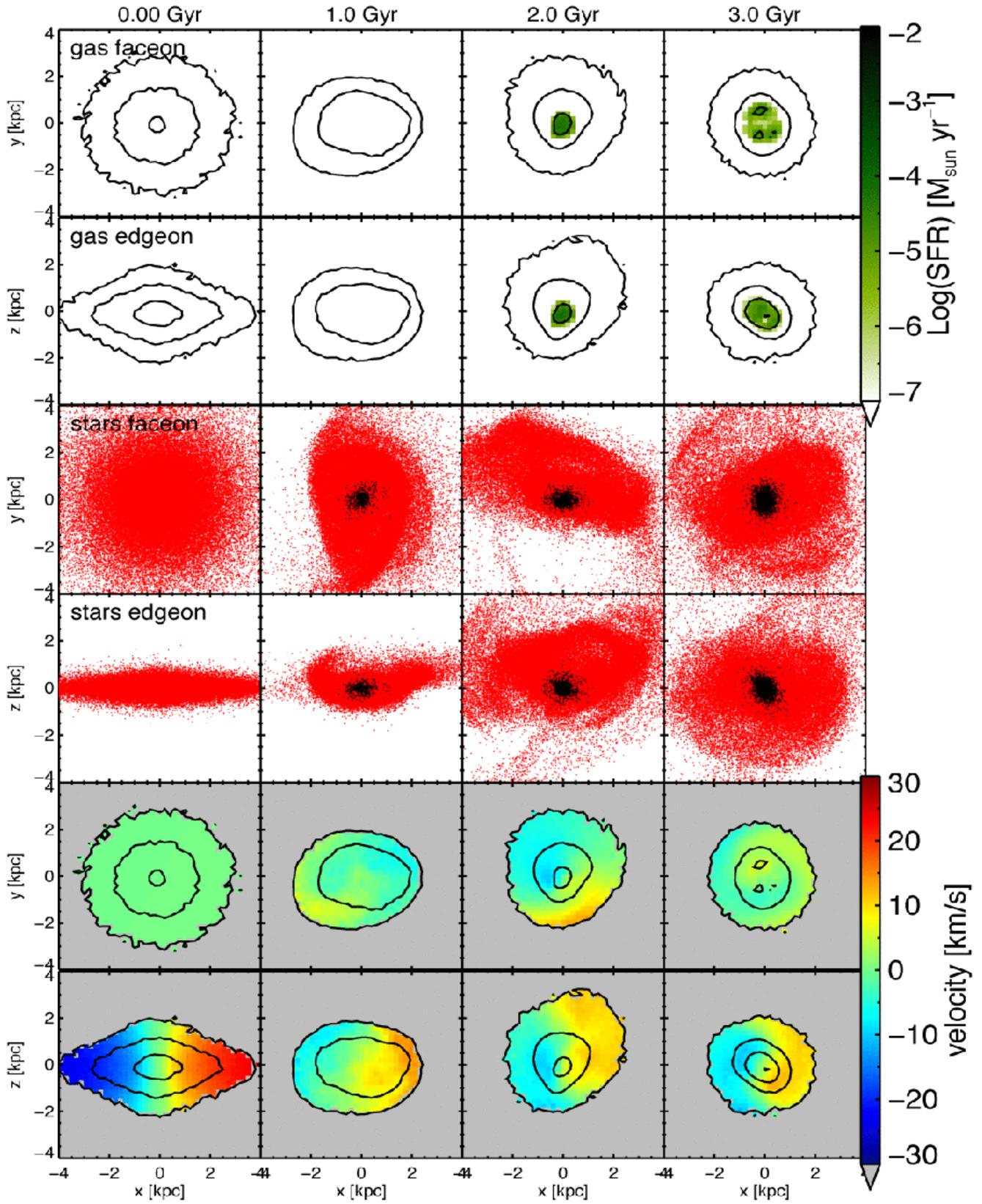}
\caption{Same as Fig.~\ref{merger-radial-highC} for the \emph{Fornax-analog} dwarf galaxy with $f_g= 0.5$ (model E1), merging with the 20\% mass satellite on a 30-degrees, radial orbit.}
\label{fnx-fg0.5}
\end{figure*}

\end{appendix}

\end{document}